\documentclass[aps,prb,twocolumn,superscriptaddress,floatfix,longbibliography]{revtex4-2}
\usepackage[utf8]{inputenc}
\usepackage[english]{babel}
\usepackage{amsmath}
\usepackage{amsfonts}
\usepackage{amssymb}
\usepackage{graphicx}
\usepackage{bm}
\usepackage{hyperref}
\usepackage[dvipsnames]{xcolor}
\usepackage[normalem]{ulem}
\usepackage{natbib}
\usepackage[percent]{overpic}
\usepackage{bbold}

\newcommand{\vect}[1]{\boldsymbol{\mathbf{#1}}}
\DeclareMathOperator{\Tr}{Tr}
\makeatletter
\DeclareRobustCommand{\cev}[1]{%
  {\mathpalette\do@cev{#1}}%
}
\newcommand{\do@cev}[2]{%
  \vbox{\offinterlineskip
    \sbox\z@{$\m@th#1 x$}%
    \ialign{##\cr
      \hidewidth\reflectbox{$\m@th#1\vec{}\mkern4mu$}\hidewidth\cr
      \noalign{\kern-\ht\z@}
      $\m@th#1#2$\cr
    }%
  }%
}

\begin{document}
\author{J. Lang}
\email[]{j.lang@uni-koeln.de}
\affiliation{Institut f\"ur Theoretische Physik, Universit\"at zu K\"oln, Z\"ulpicher Stra{\ss}e 77, 50937 Cologne, Germany}

\author{M. Buchhold}
\affiliation{Institut f\"ur Theoretische Physik, Universit\"at zu K\"oln, Z\"ulpicher Stra{\ss}e 77, 50937 Cologne, Germany}

\author{S. Diehl}
\affiliation{Institut f\"ur Theoretische Physik, Universit\"at zu K\"oln, Z\"ulpicher Stra{\ss}e 77, 50937 Cologne, Germany}

\title{Field theory for the dynamics of the open $O(N)$ model}
\date{\today}

%%
%% ABSTRACT
%%
\begin{abstract}
A field theory approach for the nonequilibrium relaxation dynamics in open systems at late times is developed. In the absence of conservation laws, all excitations are subject to dissipation. Nevertheless, ordered stationary states satisfy Goldstone's theorem. It implies a vanishing damping rate at small momenta, which in turn allows for competition between environment-induced dissipation and thermalization due to collisions. We derive the dynamic theory in the symmetry-broken phase of an $O(N)$-symmetric field theory based on an expansion of the two-particle irreducible (2PI) effective action to next-to-leading order in $1/N$ and highlight the analogies and differences to the corresponding theory for closed systems. A central result of this approach is the systematic derivation of an \emph{open-system Boltzmann equation}, which takes a very different form from its closed-system counterpart due to the absence of well-defined quasiparticles. 
%Due to the general structure of its derivation, it applies also to other open, gapless field theories under certain, testable conditions. 
%It applies generically to open, gapless field theories satisfying certain, testable conditions. 
As a consequence of the general structure of its derivation, it applies to open, gapless field theories that satisfy certain testable conditions, which we identify here. 
Specifically for the $O(N)$ model, we use scaling analysis and numerical simulations to show that interactions are screened efficiently at small momenta and, therefore, the late-time evolution is effectively collisionless. This implies that fluctuations induced by a quench dissipate into the environment before they thermalize. 
Goldstone's theorem also constrains the dynamics far from equilibrium, which is used to show that the order parameter equilibrates more quickly for quenches preserving the $O(N)$ symmetry than those breaking it explicitly.
%Furthermore, the order parameter equilibrates more quickly for quenches preserving the $O(N)$ symmetry than those breaking it explicitly. 
\end{abstract}
\maketitle

%%
%% INTRODUCTION
%%
\section{Introduction}\label{sec:Intro}
In recent years, progress in pump-probe experiments has revealed long-lived states far from equilibrium in strongly correlated materials \cite{Stojchevska2014, Liu2011, Zhang2016}. Furthermore, the order parameter dynamics and low energy excitations have been found to exhibit non-thermal critical behavior in a variety of different systems \cite{Yusupov2010, Beaud2014, Laulhe2017, Shimojima2019, Mitrano2019, Vogelgesang2018, Zong2019a, Zong2019b, Kogar2020, Zhou2021, Afanasiev2019}, with slow recovery of phase coherence reported also in the absence of topological defects \cite{Lee2012, Chuang2013, delaTorre2022}. Contrary to experiments in cold atoms \cite{Navon2016, Pruefer2018, Eigen2018, Erne2018}, which are successfully described by non-equilibrium field theory for closed systems \cite{Berges_review_2016}, the order parameter and low-lying excitations in solids \emph{strongly couple} to a variety of other relevant excitations (e.g. phonons, magnons, polarons etc.). The latter form an effective thermal bath, necessitating the order parameter and its fluctuations to be treated as an \emph{open system}.

Despite the mounting experimental evidence of scaling dynamics in out-of-equilibrium correlated materials, the theoretical description of the phenomenon is still in its infancy. Most approaches to quench dynamics to date are based on order parameter dynamics in time-dependent effective potentials \cite{Kung2013, Mihailovic2013, Ross2019, Ning2020, Dolgirev2020b, Maklar2021}. Such pictures generally disregard the continuum of gapless modes, which strictly limits them to the short-time evolution; they inevitably predict exponential relaxation, except at a dynamical critical point. Disregarding the gapless modes hence lacks the possibility of the scaling dynamics that were observed in experiments and that are expected in the vicinity of a non-thermal fixed point \cite{Berges_review_2016}.

A common approach, which takes into account order parameter fluctuations within a near-thermal framework, is the so-called three-temperature model (3-TM) \cite{Beaurepaire1996, Perfetti2007, Mansart2010, Tao2013, Pankratova2022} and its generalizations \cite{Koopmans2010}. It treats electrons, phonons, and order parameter fluctuations as thermal, albeit at different temperatures. This assumption can be made plausible when collision rates are high and the coupling between subsystems is weak. However, it is generally unclear under which conditions highly non-thermal order parameter fluctuations induced by a quench would approach an effective thermal equilibrium before dissipating.

A different strategy to step away from a pure mean-field description of the order parameter dynamics recently has been taken by the inclusion of Gaussian fluctuations \cite{Dolgirev2020a, Sun2020}. In the case of spontaneous symmetry breaking the existence of gapless Goldstone modes then leads to self-similar scaling dynamics near the thermal fixed point. However, the role of collisions between long-lived excitations has remained unexplored with no unbiased discussion of internal equilibration due to scattering and equilibration as a result of a thermal background reported in the literature.

Here we are going to close that gap by developing the dynamic theory of an $O(N)$-symmetric field theory in the spontaneously symmetry-broken phase by a two-particle irreducible (2PI) expansion to next-to-leading order in the control parameter $1/N$ (NLO(1/N)). The approach follows a similar strategy as has previously proven successful in closed systems \cite{Berges_review_2016}, with important modifications to account for the open nature of the system. This technical development culminates in the formulation of the open-system Boltzmann equation, which applies in the absence of proper quasiparticles. Although our calculations are performed on the $O(N)$ model, the arguments on the validity of the applied approximations directly translate to other models.

As a result, we obtain testable conditions on the applicability of key approximations. Specifically for quenches of the $O(N)$-symmetric theory in the symmetry-broken phase, this systematic way of proceeding allows us to show that strong screening leads to effectively collisionless dynamics. In particular, this implies that redistribution and thus thermalization is slow compared to equilibration with the environment. Consequently, excess fluctuations induced by a quench remain non-thermal until they dissipate.

We proceed as follows. We discuss an intuitive picture of the origin of slow dynamics in the symmetry-broken phase in Sec.~\ref{sec:Intuition}. We then provide a guide on how to derive the open-system Boltzmann equation starting from the microscopic model in Sec.~\ref{sec:general_theory} -- we emphasize the simplifications made and discuss their structure and necessary conditions. We start providing an in-depth discussion of our work in Sec.~\ref{sec:Goldstone}, where we introduce the 2PI effective action as a suitable foundation for our analysis. The conditions for the dynamics to be captured by a memoryless description are derived in Sec.~\ref{sec:Wigner}. In each section, emphasis is put on a physical motivation followed by detailed calculations in the later subsections. Application of this general framework in its minimal form yields the collisionless approximation analyzed in Sec.~\ref{sec:MF}. The stability of its results against collisions leading to redistribution is investigated in Sec.~\ref{sec:Collisions}. To complete the parallel with the theory of closed system dynamics, the conditions for a formulation in terms of a Boltzmann equation are derived in Sec.~\ref{sec:Boltzmann}. We conclude with an outlook concerning the applicability of our results to other systems in Sec.~\ref{sec:Outlook}

\section{Slow thermalization of quenched open systems}\label{sec:Intuition}
\emph{Physical setup ---}
We consider the evolution of open many-body systems, which at low frequencies are described by an $O(N)$-symmetric field theory with different values of $N$ and no conserved quantities. This covers a large class of condensed matter systems, like superconductors, charge-density-wave compounds, or antiferromagnets \cite{Haldane1983, Fisher1989, Zacher1998, Sachdev_book}. Following a strong quench, the order is destroyed. Since in most cases $N$ is no larger than the dimension, topological defects can be created during the quench, leading to a slow recovery of the order parameter following the Kibble-Zurek mechanism \cite{Kibble1976, Zurek1985}. Here instead, we will discuss the late-time dynamics, where either a homogeneous order has already been restored or the quench was too weak to induce defects in the first place. 
%One might falsely suppose this case to be trivial in open systems. However, as the following qualitative discussion shows, this is by no means the case. 
As the following qualitative discussion shows, this case is by no means trivial since a quick relaxation is precluded by the existence of overdamped but gapless Goldstone modes. Their dynamical mass and collisions provide two competing mechanisms, both leading to an algebraic recovery of the thermal state.

\emph{Minimal model ---}
Focusing on late times, we can drop terms, which are irrelevant in the renormalization group sense (RG-irrelevant). Then the dynamics in the $O(N)$-model is given by the Langevin equation for the real $N$-component order-parameter field $\vect{\varphi}$
\begin{align}\label{eq:model}
    \left(\partial_t^2+\gamma\partial_t-D\nabla^2+m+\frac{\lambda}{6N}\vect{\varphi}^2\right)\vect{\varphi}+\sqrt{\gamma}\vect{\xi}=0\,.
\end{align}
Here $\gamma$ parametrizes the coupling to a thermal bath at temperature $T$ (we set $\hbar=k_B=1$), which is composed of the environmental modes (e.g., phonons, etc.). The fluctuations imprinted on the order parameter by the environment are covered by the fluctuation-dissipation relation of the noise field $\vect{\xi}$
\begin{align}
    \langle \xi_a(t)\xi_b(t')\rangle=\pi T^2/\sinh^{2}{(\pi T(t-t'))}\to 2 T \delta(t-t')\, .
\end{align}
The final limit is justified if we consider the evolution of $\vect{\varphi}$ on time scales much larger than $\beta=1/T$. Under these conditions the memory of the bath with a typical correlation time or order $1/T$ can be neglected, leading to an effective model with white noise.
In the overdamped regime ($\gamma\gg \sqrt{D}|\vect{k}|$), the term $\sim \partial_t^2$ in Eq.~\eqref{eq:model} can be dropped, such that this equation corresponds to the model in \cite{Dolgirev2020a}, there studied with fluctuations treated in Hartree-Fock approximation. In the limit $\gamma\to 0$, on the other hand, with the inertial term $\sim \partial_t^2$ taken into account, it reduces to the closed system discussed in \cite{Berges_review_2016}.
We note that the crossover between these two cases takes place at short time and length scales $t\sim |\vect{r}|/\sqrt{D}\sim\gamma^{-1}$. For the late-time evolution studied here, the open character always prevails. At these late times, there is hence no smooth crossover but a qualitative difference in the time evolution for $\gamma\neq 0$ and generic interacting closed systems. This difference is made evident by considering the stationary state approached by the dynamics. While in a closed system, the initial energy density fully determines the final (thermal) distribution, the open system exchanges particles and energy with its environment. It therefore must equilibrate to a thermal distribution at the temperature of the bath and satisfy the fluctuation-dissipation relation
\begin{align}\label{eq:FDR}
\begin{split}
    i G^K(\omega,\vect{k})\equiv& \langle\vect{\varphi}(\omega,\vect{k})\vect{\varphi}(-\omega,-\vect{k})\rangle\\=&\coth{\frac{\omega}{2T}}\frac{1}{2\sqrt{\gamma}}\\ &\times\left(\frac{\delta}{\delta \vect{\xi}(-\omega,\vect{k})}\langle\vect{\varphi}(\omega,\vect{k})\rangle-\left(\omega\leftrightarrow -\omega\right)\right)\,.
\end{split}
\end{align}
This is a consequence of pump-probe experiments breaking detailed balance only by the initial quench, which results in much stronger constraints on the Langevin equation at late times than in driven open systems \cite{Zelle2023, Fruchart2021}.

\begin{figure}[th]
\centering
\includegraphics[width=.9\columnwidth]{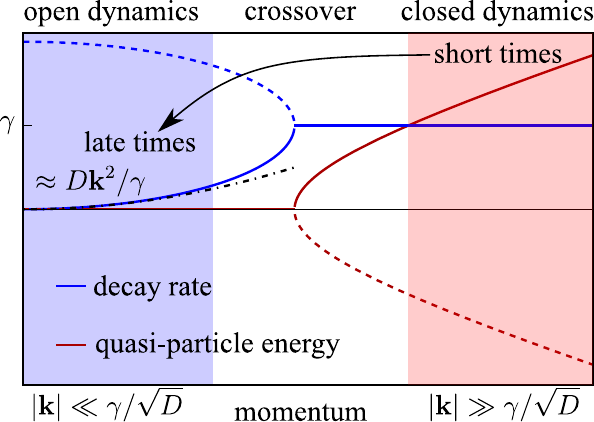}
\caption{Decay rate and excitation energy of the Goldstone mode of the open $O(N)$ model. At short times excitations at high momenta are well described by the closed system. However, the finite decay rate at high momenta at times $t\gg 1/\gamma$ only leaves long-lived, overdamped excitations at small momenta. The resulting slow dynamics is qualitatively different from that of the closed system, as is evident from the change of the dynamical exponent from $z=1$ to $z=2$.}
\label{fig:decay}
\end{figure}

For sufficiently negative values of the bare mass $m$ \footnote{Note, that fluctuations increase the effective mass compared to its bare value.} and at dimensions above the lower critical one $d=2$, the system spontaneously breaks the $O(N)$ symmetry: it acquires a nonzero expectation value for the order parameter $\phi$. Without loss of generality, we choose $\langle \varphi_\alpha\rangle=\delta_{\alpha 1}\phi$. Goldstone's theorem still holds out of equilibrium. It states that in the stationary state, interactions lead to effectively massless excitations perpendicular to the order parameter \cite{Goldstone1961}, which also affects the relaxation towards this state. The Fourier transform of the linear response function of the transversal modes (with $1<\alpha\leq N$)
\begin{align}\label{eq:GR}
\begin{split}
    G^R_\perp(t-t',\vect{r}-\vect{r}')&=-\frac{1}{2\sqrt{\gamma}}\frac{\delta}{\delta\xi_\alpha(t',\vect{r}')}\langle\varphi_\alpha(t,\vect{r})\rangle\\&\sim\theta(t-t')
\end{split}
\end{align}
in a mean-field description has two complex poles shown in Fig.~\ref{fig:decay}. The poles reveal a separation of the dynamics into two different momentum regimes: At high momenta, the dispersion is linear with a constant decay rate, $\omega\propto \pm|\vect{k}|+i\gamma$. At long wavelengths, however, fluctuations become overdamped, and the real part of the dispersion, describing reversible dynamics, vanishes identically. This allows the generation of excitations without energy cost. Focusing only on the slowly decaying mode with a vanishing decay rate $\approx D|\vect{k}|^2/\gamma$, the response function for long wavelengths takes the simple form
\begin{align}\label{eq:GRqp}
    G_\perp(\omega,\vect{k})=\frac{Z}{\omega+iD\vect{k}^2}
\end{align}
with quasiparticle weight $Z$. We, therefore, expect high-momentum excitations to decay on a time scale $t\sim 1/\gamma$ followed by algebraic relaxation due to long-wavelength fluctuations. 

On top of this single-particle picture of relaxation, one needs to consider the effect of interactions. They describe collisions, which cause a redistribution of excitations and pose an additional source for thermalization: typical states generated by a sudden quench involve an excess of fluctuations at high momenta. These scatter to lower momenta, thereby restoring a thermal distribution. If the growth rate for extremely long wavelengths due to relaxation from higher momenta is larger than the (possibly renormalized) decay rate, excitations will accumulate at small momenta, leading to an effectively even slower equilibration with modified exponents governing the algebraic decay.
To check if the quenched open $O(N)$ model realizes this scenario, we need to treat the decay rate and collisions on equal footing.

We emphasize that the Goldstone modes persist, despite the absence of any conserved quantity, due to the exchange of excitations with finite energy and momentum between the system and its environment. On the one hand, this implies the algebraic relaxation discussed above. On the other hand, on a technical level it requires approximations that introduce no new scales. These might either enter as a gap in Eq.~\eqref{eq:GRqp} or as an additional term breaking conservation laws. For example, collisions that change the particle number effectively describe two-body loss or gain. They will thus generically violate the fluctuation-dissipation relation Eq.~\eqref{eq:FDR} \cite{Sieberer2016} and hence introduce a time scale on which the deviation from the actual equilibrium begins to dominate the evolution.

\section{A systematic approach to time evolution}\label{sec:general_theory}

\begin{figure}[ht]
\centering
\includegraphics[width=\columnwidth]{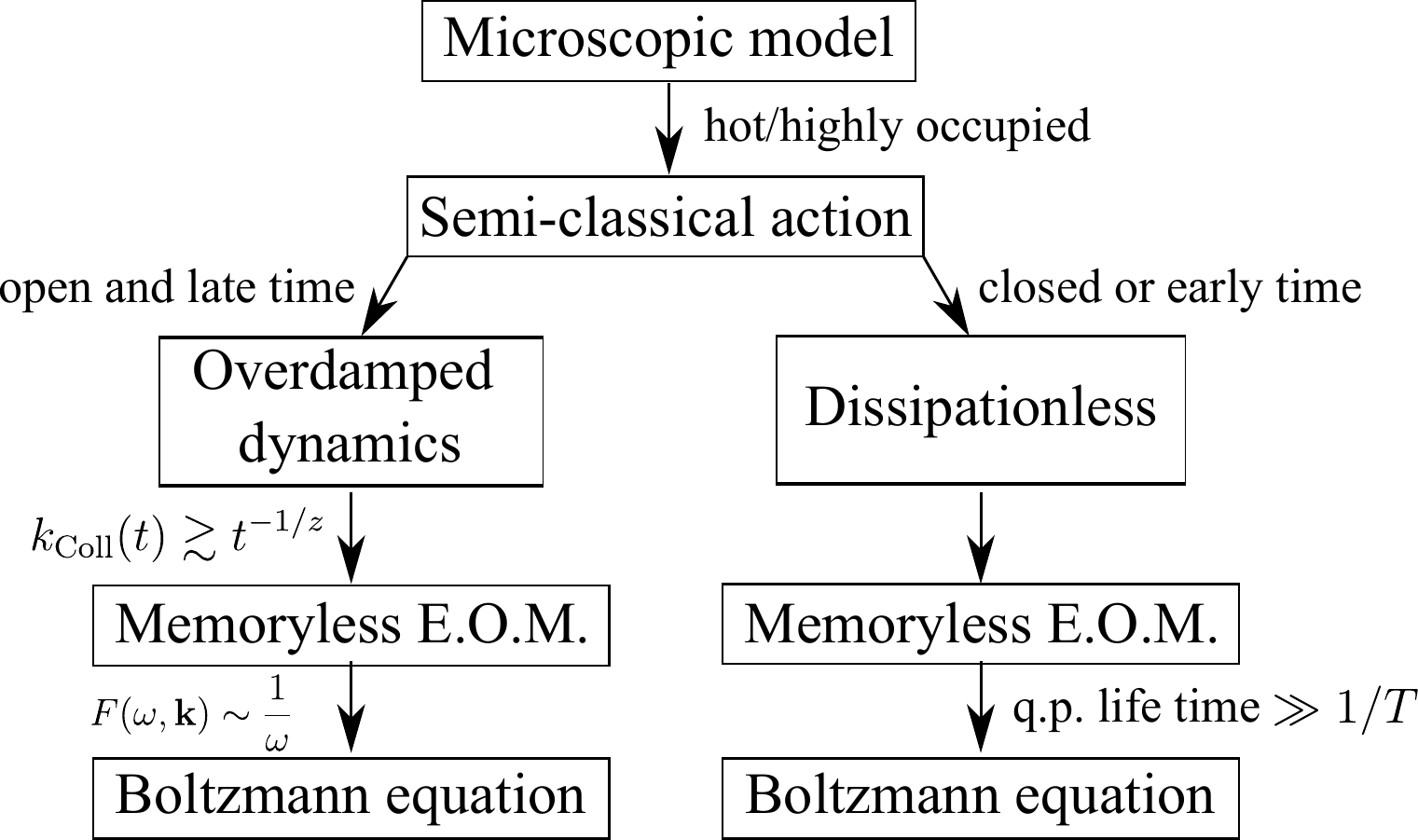}
\caption{The theory of open system dynamics is developed along the same lines as that for closed systems. While the semi-classical limit of high occupations is the same for both cases, gapless dissipative modes in open systems require a separate treatment from closed systems for all further steps towards a simple, physically insightful theory of the time evolution, see Sec.~\ref{sec:general_theory}. This schematic can be taken as a guide to this paper: Sec.~\ref{sec:Goldstone} deals with the semi-classical action. Sec.~\ref{sec:SingleMode} and Sec.~\ref{sec:Gapless} explain the steps to the memoryless equations of motion. Although numerically validated in some cases, there is no simple condition for this step in closed systems. A projection to an open-system Boltzmann equation is discussed in Sec.~\ref{sec:Boltzmann}.}
\label{fig:overview}
\end{figure}

With the physical picture from the previous section in mind, we now state the general conditions for a faithful theory of the evolution of gapless open systems:
\begin{itemize}
    \item Once a non-zero order parameter breaks the symmetry and the excitations become gapless, interactions can no longer be considered to be weak: in the absence of scales a perturbative expansion in the interaction strength is no longer controlled. Truncating such an expansion at a finite order would inevitably introduce a bias between intrinsic collisions and particle or energy exchange with the bath. Instead, one needs an ordering principle that applies to strongly interacting systems. Here we will use an expansion in the inverse number of components of the order parameter field $1/N$. This expansion has proven surprisingly accurate already for moderate values of $N$ \cite{Berges2003}. It does however fail to resolve topological defects, which can be generated in systems with $N\leq d$ and for strong quenches as the order parameter has to vanish locally \cite{Bray2002}. Hence, for the weak quenches considered here, they can be safely neglected.
    \item When the decay of excess fluctuations at late times becomes algebraically slow, it is vital that none of the used approximations introduces an additional, unphysical time scale for relaxation. One way to guarantee this is through the use of conserving approximations \cite{Baym1962}, which satisfy the same conservation laws as the microscopic model. Although the open $O(N)$ model has no conserved quantities, their absence is a result of the coupling to a thermal bath. Collisions, however, conserve energy and particle number. A non-conserving approximation could introduce a time scale beyond which this is violated and qualitatively change the approach to equilibrium. The use of a conserving approximation furthermore has the advantage that our approach easily generalizes to models with conserved quantities \footnote{Although irrelevant for weak quenches, conserving interactions are known to exhibit self-organized criticality in the Olami-Feder-Christensen model \cite{Olami1992}. This is not true for non-conserving interactions \cite{Grassberger1994,Carvalho2000,Boulter2003,Wissel2006}, which exemplifies the non-trivial role conservation laws play in local collisions, even in the absence of conserved quantities.}.
    \item The approximation must not break the $O(N)$-symmetry explicitly. It needs to satisfy Goldstone's theorem, necessary for the algebraic slow-down of the relaxation at late times. This condition, which relates to the spectrum of the model, has to be viewed independently of the previous one concerning the stationary distribution.
    \item For practical purposes it is highly beneficial if the evolution equations can be simulated with an effort that scales at most linear in time.
\end{itemize}

In the following, we develop a theory for the late-time evolution in the spontaneously symmetry-broken phase. To this end, we systematically eliminate processes that are irrelevant at long times and at large distances, which leads to simplifications in the evolution equations. Although this process follows the same general logic as the corresponding theory for closed systems, the performed approximations and the resulting equations contain crucial differences. The similarities and differences in the approach to closed or open systems are visualized in Fig.~\ref{fig:overview}. 

A major consequence of combining gapless modes and thermal fluctuations is that the microscopic model at small momenta can be reduced to the semi-classical one stated in Eq.~\eqref{eq:model}, which describes the evolution of open systems at late times. We stress in Sec.~\ref{sec:SingleMode}, that for closed systems or open systems at early times, a different semi-classical evolution emerges, that cannot be connected to the late time evolution of open systems by a smooth crossover. As a result, the subsequent simplifications require different arguments for open systems at late times compared to closed systems or early times.

Along these lines, we demonstrate an intriguing and useful feature of open, gapless systems: if collisions are dominated by a time-dependent momentum scale $k_\text{Coll}(t)\gtrsim t^{-1/z}$ with $z$ the dynamical exponent, the overdamped evolution will no longer depend on the entire history of correlation functions. In other words, the evolution becomes Markovian (memoryless), such that interactions -- to a good approximation -- include only correlation functions that are local in time. In particular, it implies that the system evolution becomes independent of the details of the initial quench. This drastically simplifies any numerical treatment and is a prerequisite of scaling dynamics. 

Although the system is coupled to a bath modeled as uncorrelated, white noise, the Markovian evolution is not a direct consequence of the latter. Instead, it results from a slow evolution compared to the lifetime of the participating excitations. As such it is sensitive to the rate of collisions and the distribution of long-lived modes.
Such an effective Markovian evolution has also been confirmed numerically for some observables in closed systems \cite{Berges2006}. There exists, however, no rigorous justification for this behavior in closed systems. 

The final result of our strategy will be the derivation of an \emph{open-system Boltzmann equation}. Due to the absence of well-defined quasiparticles, this Boltzmann equation takes a very different form compared to its closed system counterpart. The validity of a Boltzmann approximation in open systems requires a $1/\omega$ divergence in the infrared distribution function $F(\omega,\vect{k})$. This is justified when the state obeys a near-thermal long-wavelength distribution. However, we emphasize that it applies significantly more generally since the momentum dependence of $F$ remains unconstrained, corresponding to a momentum-dependent temperature $T(\vect{k})$. In fact, it should be viewed as the open-system counterpart of long-lived quasiparticles, which are the theoretical foundation for the Boltzmann approximation in closed systems.
%This is justified when the state obeys a near-thermal long-wavelength distribution.
%We emphasize that this, however, does not require a near-thermal distribution since the momentum dependence of F remains unconstrained. We contrast this with the condition for the Boltzmann approximation in closed systems, where it applies under the rather weak condition of long-lived quasiparticles

The application of the newly derived memoryless framework to the open $O(N)$-symmetric model \eqref{eq:model} reveals that collisions are heavily screened at long wavelengths by gapless excitations. As a result, momentum transfer between fluctuations plays no significant role in the thermalization at late times. Instead, the asymptotic relaxation is governed by an effective single-particle picture, and an order parameter-dependent dynamical mass. Consequently, the approach to the thermal state is described by Gaussian exponents. The behavior of the dynamical mass is further constrained by Goldstone's theorem. This imprints a subthermal momentum distribution of the excess fluctuations for all quench protocols that fulfill our initial conditions, i.e., that preserve the $O(N)$ symmetry. Contrary to the scenario, where the $O(N)$ symmetry is explicitly broken at early times, the order parameter fluctuations will not thermalize to an effective temperature different from that of the environment. Instead, the long wavelength fluctuations retain their subthermal distribution until they decay into the bath. This process leads to a thermal distribution with the temperature of the bath. Absent significant redistribution no intermediate regime of different temperatures for the order parameter fluctuations and the environment exists at late times. This is in contrast to the general assumption underlying the three temperature models. The latter instead qualitatively overestimate the slowing down of the order parameter relaxation.

%\section{Goldstone mass}\label{sec:Goldstone}
\section{Conserving and gapless effective actions}\label{sec:Goldstone}
Our goal is to construct a controlled, non-perturbative expansion and thus we turn to the field-theoretic description of Eq.~\eqref{eq:model}. This gives access to the tools of non-equilibrium field theory, which facilitates a systematic formulation. The corresponding Martin-Siggia-Rose-Janssen-DeDominicis (MSRJD) action \cite{Martin1973a,Janssen1976a,DeDominicis1976} can be obtained from the full quantum mechanical description of the system at low frequencies, where the large amount of thermal fluctuations allows one to neglect quantum fluctuations $\coth{\omega/(2T)}\gg 1$. This justifies the classical description employed in Eq.~\eqref{eq:model} \cite{Kamenev_book,Sieberer2016}. As a technical side note, we point out that in the dissipationless case, the classical action describes a deterministic evolution of the field $\varphi_\text{cl}$ \cite{Aarts2002}, which can be simulated efficiently \cite{Orioli2015}. For the open system, however, even in the absence of quantum fluctuations, thermal fluctuations from the bath significantly increase the numerical cost of such simulations.
%\jlc{I think it is useful to mention that the Langevin equation is harder to simulate than the closed system, which removes the main alternative to our approach.}

In the following, we are using Einstein notation, such that the MSRJD action takes the form
\begin{align}\label{eq:action}
\begin{split}
    S[\varphi]&=S_0[\varphi]+S_\text{int}[\varphi]\\
    S_0[\varphi]&=\frac{1}{2}\!\int\!\! dt\!\int\!\! d^3 r
        \begin{pmatrix}
        \varphi_{\alpha,\text{cl}} & \varphi_{\alpha,\text{q}}
        \end{pmatrix}
        \mathcal{G}_0^{-1}(t,\vect{r})
        \begin{pmatrix}
        \varphi_{\alpha,\text{cl}} \\ \varphi_{\alpha,\text{q}}
        \end{pmatrix}\\
    S_\text{int}[\varphi]&=-\frac{\lambda}{3N}\!\int\! dt\!\int\! d^3 r\, \varphi_{\alpha,\text{q}} \varphi_{\alpha,\text{cl}} \varphi_{\beta,\text{cl}} \varphi_{\beta,\text{cl}} %+ \varphi_{\alpha,\text{cl}} \varphi_{\alpha,\text{q}} \varphi_{\beta,\text{q}} \varphi_{\beta,\text{q}}
\end{split}
\end{align}
with the matrix-valued inverse Green's function
\begin{align}\label{eq:invG0}
\begin{split}
        \mathcal{G}_0^{-1}(t,\vect{r})&=\begin{pmatrix}
        0 & \left[\mathcal{G}^A_0\right]^{-1}(t,\vect{r})\\
        \left[\mathcal{G}^R_0\right]^{-1}(t,\vect{r}) & 8i T \gamma
        \end{pmatrix}\\
        \left[\mathcal{G}^A_{0,\alpha\beta}\right]^{-1}(t,\vect{r})&=2\left(-{\partial_t^A}^2-\gamma \partial_t^A+D \vect{\nabla}^2_r-m\right)\delta_{\alpha\beta}\\
        \left[\mathcal{G}^R_{0,\alpha\beta}\right]^{-1}(t,\vect{r})&=2\left(-{\partial_t^R}^2-\gamma \partial_t^R+D \vect{\nabla}^2_r-m\right)\delta_{\alpha\beta}\,.
\end{split}
\end{align}
Here $\partial_t^{R/A}$ denote the retarded/advanced time derivatives and $\mathcal{G}_0^{R/A}$ are the bare retarded/advanced Green's functions. They are diagonal in the space of field components, which we label by Greek indices.

In the symmetry-broken phase, it is convenient to directly expand in fluctuations around the proper saddle-point $\langle \varphi_\alpha\rangle=\delta_{\alpha 1}\phi$. We therefore define the generating function $\mathcal{Z}[J,R]$ as the partition function of a non-equilibrium system in the presence of source fields $J(x)$ and $R(x,y)$. Here and in the following, we use the shorthand notation $x=(t,\vect{r})$. One then finds \cite{Berges_review_2016,Kamenev_book}
\begin{align}
\begin{split}
    \mathcal{Z}[J,R]&\equiv \exp{(iW[J,R])}\\&=\!\int\! D\varphi \exp i\left\{S[\varphi]+\int_{x}J_{\alpha,\bar{a}}(x)\varphi_{\alpha,a}(x)\right.\\ &\qquad\quad\left.+\frac{1}{2}\!\int_{x,y}\!\varphi_{\alpha,a}(x)R_{\alpha\beta,\bar{a}\bar{b}}(x,y)\varphi_{\beta,b}(y)\!\right\},
\end{split}
\end{align}
where Roman indices run over Keldysh space $(\text{cl},\text{q})$ and $\bar{a}$ represents the complement of $a$. All properties of the system can be obtained by taking functional derivatives of $\mathcal{Z}[J,R]$ or the Legendre transformations of $W[J,R]$. The latter is often more convenient due to the direct dependence on physical quantities, namely the order parameter expectation value and correlation functions.

For example, the one-particle irreducible (1PI) action $\Gamma[\phi]$ is obtained by a Legendre transformation with respect to the linear source $J$ \cite{Schwinger1951a, Schwinger1951b, Schwinger1958, DeDominicis1963, DeDominincis1964, Jona-Lasinio1964}
\begin{align}\label{eq:1PI-action}
\begin{split}
    \Gamma[\phi]&=W[J,0]-\int_{x} J_{\alpha,a}(x)\phi_{\alpha,\bar{a}}(x)\\&=S[\phi]+\frac{i}{2}\Tr_\mathcal{C}\ln{[G_0^{-1}(\phi)]}+\Gamma_1[\phi]\,
\end{split}
\end{align}
where $S[\phi]$ represents the saddle-point action, i.e. the action evaluated at the classical field $\vect{\phi}=\langle\vect{\varphi}\rangle$, and the $\Tr_\mathcal{C}$ corresponds to an integration over space-time together with a summation over field components and a trace in Keldysh space. The bare, inverse propagator is given by
\begin{align}
    \left[G_{0}^{-1}(x,y,\phi)\right]^{a b}_{\alpha\beta}=\frac{\delta^2 S[\phi]}{\delta \phi_{\alpha,\bar{a}}(x)\delta\phi_{\beta,\bar{b}}(y)}\,.
\end{align}
Since $\Gamma_1[\phi]$ contains all one-particle irreducible diagrams,
\begin{align}
    \left[G^{-1}(x,y,\phi)\right]^{a b}_{\alpha\beta}=\frac{\delta^2\Gamma[\phi]}{\delta \phi_{\alpha,\bar{a}}(x)\delta\phi_{\beta,\bar{b}}(y)}
\end{align}
is the inverse of the connected two-point function (see for example \cite{Amit1984}).

We notice, that the effective action achieves our goal of expanding around the classical field $\phi$. It thus expands around the true vacuum and has the advantage of naturally capturing the symmetry-broken phase $\phi\neq 0$, where it additionally allows the derivation of Ward-Takahashi identities \cite{Ward1950, Takahashi1957} and thus the Goldstone theorem \cite{Goldstone1961}. However, it turns out that theories derived from $\Gamma[\phi]$ are not conserving, meaning that independently of the coupling to the bath, collisions alone will violate macroscopic conservation laws. This artifact of the approximation is elaborated on in App.~\ref{sec:1PI}. Intuitively, one may say that a conserving approximation needs to derive from a functional that evolves the order parameter and its fluctuations with a consistent set of equations. Particles removed from one part of the system by collisions must consistently be added elsewhere. It is therefore not allowed to refer to any particular, inert state, which requires an expansion in $G$ instead of $G_0$ in the interacting part of the effective action \cite{Kadanoff_book}. This can be achieved by a second Legendre transformation with respect to the two-point source $R(x,y)$, which leads to the two-particle irreducible (2PI) action \eqref{eq:2PI-action}. This moreover has the pleasant, purely technical advantage of significantly reducing the number of Feynman diagrams that need to be evaluated.

The result of this procedure is the non-equilibrium equivalent of the Luttinger-Ward functional \cite{Luttinger1960} known as 2PI effective action \cite{Baym1962, Cornwall1974}
\begin{align}\label{eq:2PI-action}
\begin{split}
    \Gamma[\phi,G]&=S[\phi]+\frac{i}{2}\Tr_\mathcal{C}\ln{G^{-1}}+\frac{i}{2}\Tr_\mathcal{C}\left(G_0^{-1}(\phi)G\right)\\&\quad+\Gamma_2[\phi,G]\,.
\end{split}
\end{align}
Here $\Gamma_2[\phi,G]$ includes all two-particle irreducible diagrams written in terms of the dressed propagator
\begin{align}\label{eq:Dyson}
    G_{\alpha\beta}(x,y)=\left[G_{0}^{-1}(x,y)-\Sigma(x,y,\phi,G)\right]^{-1}_{\alpha\beta}\,,
\end{align}
where the proper self-energy
\begin{align}
    \Sigma_{\alpha\beta}^{a b}(x,y,\phi,G)=2i\frac{\delta\Gamma_2[\phi,G]}{\delta G^{\bar{a}\bar{ b}}_{\alpha\beta}(x,y)}
\end{align}
involves only one-particle irreducible diagrams. In the last equation, we have made the Keldysh structure explicit with Keldysh index $(a, b)\in\{\text{cl},\text{q}\}$ and the identification $G^{\text{q},\text{cl}}=G^R$, $G^{\text{cl},\text{q}}=G^A$, $G^{\text{cl},\text{cl}}=G^K$ and the vanishing component $G^{\text{q},\text{q}}=G^V$.

If a partition function $\mathcal{Z}$ possesses a spontaneously broken continuous symmetry, massless Goldstone modes exist to all orders in perturbation theory. This implies a divergent correlation length \cite{Goldstone1961}, which in turn results in the breakdown of perturbation theory. A systematic, non-perturbative expansion is instead obtained by truncating the infinite set of diagrams contributing to $\Gamma_{2}$ at a given order in $1/N$. This corresponds to the summation of select interactions to infinite order. It is, however, not a priori clear, whether a non-perturbative approximation to $\mathcal{Z}$ retains its global symmetry. As we have stated before, the latter is a prerequisite of the Goldstone theorem and therefore a massless transverse mode in the stationary state. Hence, we need to check if the mass of the transverse mode of the $O(N)$ model vanishes in 1PI and 2PI in next-to-leading order in $1/N$ (NLO($1/N$)).

We show in Sec.~\ref{sec:1PI} that the masslessness of the transverse mode is preserved for the 1PI effective action, but not the 2PI effective action. The latter is, however, required for a conserving approximation, see Sec.~\ref{sec:2PI}. This is a well-known, long-standing problem for conserving approximations \cite{Baym1977, Haussmann2007}. In general, one finds that in 2PI the mass only vanishes up to the order of the expansion, i.e. for an expansion to NLO($1/N$) one finds a Goldstone mass $\sim N^{-2}$. In the following we will avoid this problem by using a low-energy quasiparticle parametrization of the transverse Green's function $G^R_\perp$ as in Eq.~\eqref{eq:GRqp} that is explicitly massless.

\subsection{1PI effective action}\label{sec:1PI}
The $1/N$ expansion of the 1PI effective action truncates all equations of motion in the same order in $1/N$ and therefore exactly preserves the symmetry of the $O(N)$ model. Although this statement is more general, we show it explicitly up to next-to-leading order in $1/N$. To make the discussion more tangible, we will provide the results here, relegating the details of their derivation to App.~\ref{app:1PI}. 

At leading order in the expansion in $1/N$ the equation of motion for the order parameter field is obtained from the functional derivative
\begin{align}\label{eq:1PI_op}
\begin{split}
    \frac{\delta \Gamma^\text{LO}}{\delta\phi_\text{q}(x)}\bigg|_{\phi=\langle\hat{\phi}\rangle}\!\!=&\bigg[-\partial_t^2-\gamma\partial_t-m+D \vect{\nabla}_r^2\\&\;\left.-\frac{\lambda}{6N}\!\left(\phi^2(x)+i G^K_\text{LO}(x,x,\phi)\right)\!\right]\!\phi(x)\!\overset{!}{=}\!0\,.
\end{split}
\end{align}
Here $G_{\text{LO}}$ denotes the full Green's function in Hartree approximation \footnote{Note, that this is not the same as the 2PI Green's function at leading order in $1/N$} given by
\begin{align}\label{eq:GHartree}
\begin{split}
    G_{\text{LO},\perp/\parallel}^{-1}(x,y,\phi)=&G_{0,\perp/\parallel}^{-1}(x,y,\phi)\\&-\frac{i\lambda}{3N}\sigma^1\delta_{\alpha\beta} \delta^{(4)}(x-y)G^K_\text{LO}(x,x,\phi)\,,
\end{split}
\end{align}
where the first Pauli matrix $\sigma^1$ acts in Keldysh space and the momentum dependence has been suppressed. To ease the physical interpretation, we have made the distinction between transversal ($\perp$) and longitudinal ($\parallel$) modes explicit. Functions without this label are to be summed over all modes, i.e. $N-1$ transversal and one longitudinal mode. Consequently, Eq.~\eqref{eq:GHartree} is to be solved self-consistently for $G_{\text{LO},\perp/\parallel}$. We note that the Hartree approximation describes a dynamical shift of the bare mass proportional to the density of fluctuations $n(x)=i/2 G^K_{LO}(x,x,\phi)$. Even before the inclusion of any loop corrections, in the symmetry-broken phase, the retarded and advanced Green's functions are shifted by the order parameter field
\begin{align}\label{eq:Gphi}
\begin{split}
    \left[G_{0,\perp}^{-1}(x,y,\phi)\right]^{R/A}=&\left[\mathcal{G}_{0,\perp}^{-1}(x,y)\right]^{R/A}\\&-\frac{\lambda}{3N}\delta^{(4)}(x-y)\phi^2(x)\\
    \left[G_{0,\parallel}^{-1}(x,y,\phi)\right]^{R/A}=&\left[\mathcal{G}_{0,\parallel}^{-1}(x,y)\right]^{R/A}\\&-\frac{\lambda}{N}\delta^{(4)}(x-y)\phi^2(x)\,.\\
\end{split}
\end{align}

Similarly to the order parameter dynamics, the Green's functions at leading order in $1/N$ are obtained from the second functional derivative of the effective action with respect to $\phi_{\text{q},\text{cl}}$. To check the validity of Goldstone's theorem it would be enough to find the retarded Green's function of the transversal mode, but for completeness, we give the full result as $G_{\perp/\parallel}^{-1}(x,y,\phi,G)=G_{0,\perp/\parallel}^{-1}(x,y,\phi,G)-\Sigma_{\perp/\parallel}(x,y,\phi,G)$, with
\begin{align}\label{eq:Sigma1PI}
\begin{split}
    \Sigma^R_\perp(x,y)=&\frac{i\lambda}{3N}\delta^{(4)}(x-y)G^K_{\text{LO}}(x,x)\\
    \Sigma^R_\parallel(x,y)=&\frac{i\lambda}{3N} \delta^{(4)}(x-y) G^K_{\text{LO}}(x,x)\\&+\phi(x)\phi(y)\left(\hat{\Pi}^R(x,y)-\frac{2\lambda}{3N}\delta^{(4)}(x-y)\right)\\
    \Sigma^K_\perp(x,y)=&0\\
%    \Sigma^K_\parallel(k)=&\frac{i}{2}|\Pi^R(k)|^2\phi^2\\&\times\int_q \left(G^K_{\text{LO}}(k-q)G^K_{\text{LO}}(q)+G^R_{\text{LO}}(k-q)G^R_{\text{LO}}(q)+G^A_{\text{LO}}(k-q)G^A_{\text{LO}}(q)\right)\,,
    \Sigma^K_\parallel(x,y)=&\frac{i}{2}\int_{z_1,z_2}\hat{\Pi}^R(x,z_1)\hat{\Pi}^A(z_2,y)\phi(x)\phi(y)\\&\!\times\!\left({G^K_{\text{LO}}}^2+{G^R_{\text{LO}}}^2+{G^A_{\text{LO}}}^2\right)\!(x,y)\,,
\end{split}
\end{align}
where
\begin{align}\label{eq:1-PI_Vertex}
    \hat{\Pi}^R(x,y)=\left(\frac{3N}{2\lambda}\delta^{(4)}(x-y)-i \left(G^K_{\text{LO}} G^R_{\text{LO}}\right)(x,y)\right)^{-1}\,.
\end{align}
The caret on $\hat{\Pi}^R$ is used to distinguish the screened vertex from the related expression that will later be derived from the 2PI effective action.

We note that all self-energy contributions are of order $N^0$, with all diagrams of this order included. This is a general property of the 1PI scheme. It implies, that the longitudinal mode can decay into two transversal modes. But because there are $N-1$ transversal and only one longitudinal mode, the corresponding term is absent in the transversal self-energy. Consequently, the approximation cannot be conserving as collisions alone violate particle number and energy conservation: longitudinal modes that decay into transverse excitations are simply lost. On the other hand, 1PI contains all diagrams to a given order in $1/N$. Thus it is effectively a perturbation theory in $1/N$ and satisfies Ward-Takahashi identities. This is easy to check in the symmetry-broken phase. The stationarity condition of the homogeneous field expectation value
\begin{align}
    m+\frac{\lambda}{6N}\left(\phi^2+i\int_q G^K_{\text{LO}}(q)\right)=0\,
\end{align}
with $G(q)=\int_x e^{-i q (x-y)}G(x-y)$ the Fourier transform of the translation invariant Green's function, implies $\left[G^R_\perp\right]^{-1}(k=0)=0$. Hence to leading order, the 1PI effective action satisfies Goldstone's theorem.

At next-to-leading order, the complexity of the 1PI calculation increases drastically. It is portrayed in App.~\eqref{app:1PI-NLO} and we only state its conclusion here: in next-to-leading order, in $1/N$ the equation of motion for the order parameter can be written as
\begin{align}\label{eq:Goldstone1PI}
    0\overset{!}{=}\frac{\delta \Gamma^\text{NLO}}{\delta \phi_\text{q}(k)}=G^R_\perp(k)\phi(k)\,,
\end{align}
from which again readily follows Goldstone's theorem.

At next-to-leading order in $1/N$, the decay of longitudinal excitations into transversal modes is balanced by a corresponding process for the longitudinal mode. There are however additional processes $\sim 1/N$ in the longitudinal self-energies that lack a reverse process for the transversal modes. From this, a general structure emerges: At any expansion order $m$ the Goldstone theorem is satisfied exactly, but conservation laws are broken by processes $\sim N^{-m-1}$. At next-to-leading order, this implies that at time scales $t\gtrsim N^{2}\tau$, with $\tau$ the collision time, unbalanced collisions can become dominant at small momenta as the equilibration rate due to the coupling to the bath vanishes there.

\subsection{2PI effective action}\label{sec:2PI}

\begin{figure}[ht]
\centering
\includegraphics[width=.5\columnwidth]{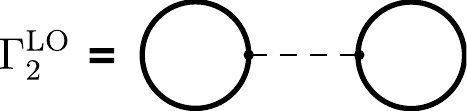}
\caption{At each order in $1/N$ 2PI requires far fewer diagrams than 1PI. This becomes apparent already at leading order, where due to the restriction to skeleton diagrams (i.e. those without self-energy insertions) only a single diagram contributes to the effective 2PI action. Bold lines indicate dressed Green's functions and the thin dashed line represents the contact interaction $\sim \frac{\lambda}{3N}\delta(x-x')$.} 
\label{fig:Gamma2LO}
\end{figure}

Aiming for a conserving approximation, we now turn our attention to the 2PI effective action, which we derive and analyze the equations of motion at the same orders in $1/N$. Fortunately, there are far fewer two-particle irreducible diagrams than terms in the corresponding 1PI effective action. Specifically at leading order, only one (bold) diagram contributes (see Fig.~\ref{fig:Gamma2LO}). It takes the simple form
\begin{align}
    \Gamma^\text{LO}_2=\frac{\lambda}{6N}\Tr_\mathcal{C}\left(G(x,x)\cdot\tau^1\cdot G(x,x)\right)\,,
\end{align}
where $\tau^1=\mathbb{1}_{N\times N}\otimes \sigma^1$ represents the identity matrix on the space of field components and the Pauli matrix $\sigma^1$ in Keldysh space.

Clearly, at this level, no collisions occur and the Green's functions are
\begin{align}\label{eq:MF_GF}
\begin{split}
    \left[G^{-1}\right]^{R/A}_{\alpha\beta}(x,y)=& \left[G_0^{-1}\right]^{R/A}_{\alpha\beta}(x,y)\\&-\frac{i\lambda}{3N}\delta^{(4)}(x-y)G^K_{\gamma\gamma}(x,x)\\
    \left[G^{-1}\right]^{K}_{\alpha\beta}(x,y)=&\left[G_0^{-1}\right]^{K}_{\alpha\beta}(x,y)\\
    \left[G^{-1}\right]^{V}_{\alpha\beta}(x,y)=&0\,,
\end{split}
\end{align}
with repeated indices being summed over and the bare inverse Green's functions again shifted by the expectation value of the order parameter $\phi$. We have also introduced the superscript $V$ to label the upper left entry of the inverse Green's function in Eq.~\ref{eq:invG0}. Although causality implies that it must vanish for every physical system, it still needs to be considered when taking functional derivatives.

As for the one-particle irreducible formulation, the field expectation value satisfies
\begin{align}
0\overset{!}{=}\frac{\delta \Gamma^\text{LO}}{\delta \phi_\text{q}(x)}\,,
\end{align}
with the subtle distinction, that $\delta G/\delta \phi=0$ since $\Gamma_2$ is a functional of both $\phi$ and $G$. Hence to leading order in $1/N$
\begin{align}\label{eq:MF_phi}
\begin{split}
    &\left(\partial_t^2+\gamma\partial_t+m-D \vect{\nabla}^2\right.\\&\quad\left.+\frac{\lambda}{6N}\left(\phi^2(x)+iG^K(x,x)\right)\right)\phi(x)=0
\end{split}
\end{align}
and the condition for a symmetry-broken state is formally the same as in 1PI. Since the same is true for the transversal Green's function $G^R_\perp$, we find that Goldstone's theorem is satisfied exactly for an expansion to leading order in $1/N$. However, contrary to there, the longitudinal self-energy contains no collision effects, which as we discussed before would lead to the violation of conservation laws.

\begin{figure}
\centering
\includegraphics[width=.8\columnwidth]{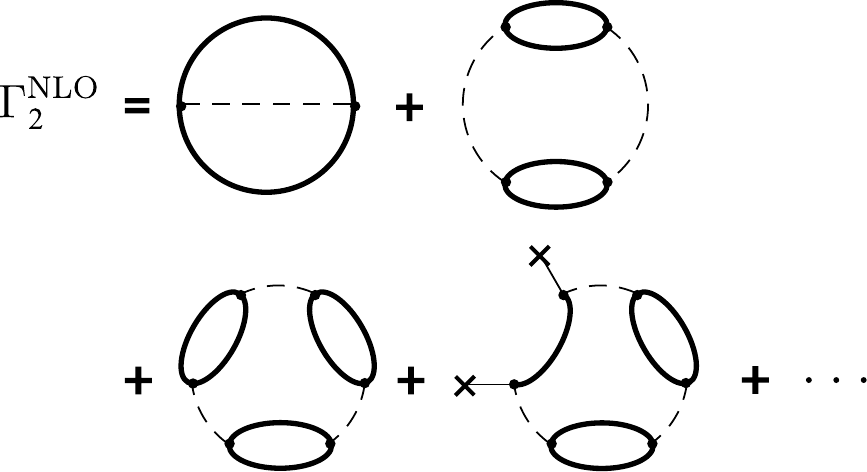}
\caption{The diagrammatic representation of $\Gamma_2$ up to NLO($1/N$) is the sum of all interaction rings with any number of screening bubbles inserted. In each diagram, one bubble may be opened with one Green's function being replaced by two fields if the cut diagram involves at least two loops. Here we show the first three terms of this series.} 
\label{fig:Gamma2NLO}
\end{figure}

Without collisions, the theory is unable to describe redistribution and thus thermalization in the absence of a bath.
We therefore must proceed to include the next-to-leading order in $1/N$. The diagrams involved are shown in Fig.~\ref{fig:Gamma2NLO}. The translation into an integral expression including the Keldysh structure can be written in the fairly compact form
\onecolumngrid
\begin{align}\label{eq:Gamma2NLO}
\begin{split}
    \Gamma_2^\text{NLO}=&\int_{x,y}\left(1-i\phi_\text{q}(x)\phi_\text{cl}(y)\left(\frac{\delta}{\delta G^R_\parallel(x,y)}+\frac{\delta}{\delta G^A_\parallel(x,y)}\right)-i\phi_\text{cl}(x)\phi_\text{cl}(y)\frac{\delta}{\delta G^K_\parallel(x,y)}\right)\Gamma_2^\text{RPA}\\&+i\frac{\lambda}{3N}\int_x\left(2\phi_\text{cl}(x)\phi_\text{q}(x)G^K_\parallel(x,x)+\phi_\text{cl}^2(x)\left(G^R_\parallel(x,x)+G^A_\parallel(x,x)\right)\right)\,,
\end{split}
\end{align}
with
\begin{align}\label{eq:Gamma2RPA}
    \Gamma_2^\text{RPA}=&\frac{i}{2}\Tr_\mathcal{C}\ln\left[\mathbb{1}-\frac{2\lambda}{3N}\tilde{V}\right]
\end{align}
and the polarization bubble
\begin{align}
    \tilde{V}_{\alpha,\beta}(x,y)=\frac{i}{2}\begin{pmatrix}
        \Tr(G_{\alpha,\beta}(x,y)\cdot G^\top_{\alpha,\beta}(x,y))&\Tr(G_{\alpha,\beta}(x,y)\cdot \sigma^1 \cdot G^\top_{\alpha,\beta}(x,y))\\
        \Tr(G_{\alpha,\beta}(x,y) \cdot G^\top_{\alpha,\beta}(x,y)\cdot \sigma^1) & \Tr(G_{\alpha,\beta}(x,y)\cdot \sigma^1 \cdot G^\top_{\alpha,\beta}(x,y)\cdot \sigma^1)
    \end{pmatrix}\,.
\end{align}
\twocolumngrid
%\begin{align}
%    \tilde{V}_{\alpha,\beta}(x,y)=\frac{i}{2}\!\!\begin{pmatrix}\!
%        \Tr(G\cdot G^\top)&\Tr(G\cdot \sigma^1 \cdot G^\top)\\
%        \Tr(G \cdot G^\top\cdot \sigma^1) & 
%        \!\Tr(G\cdot \sigma^1 \cdot G^\top\cdot\sigma^1)\!
%    \end{pmatrix}_{\!\alpha,\beta}\!\!\!\!\!(x,y)
%\end{align}
\noindent In the last expression the trace and matrix multiplication as well as the matrix act in Keldysh space. In Eq.~\eqref{eq:Gamma2RPA} $\mathbb{1}$ denotes the identity with respect to space-time, Keldysh space, and field index. The last line in Eq.~\eqref{eq:Gamma2NLO} subtracts the one-loop diagrams created by the derivatives in the first line as these contributions are already included in $\frac{i}{2}\Tr_\mathcal{C}\left(G_0^{-1}[\phi]G\right)$. 

The evaluation of the Dyson equation obtained from $\Gamma_2^\text{NLO}$ is numerically very expensive with a computational cost that scales as $\mathcal{O}(t_\text{sim}^3)$, where $t_\text{sim}$ is the simulated time. In section \ref{sec:Wigner} we therefore justify approximations that allow us to use time-local expressions that only depend on the relative coordinate $x-y$. Since these are the equations solved in Sec.~\ref{sec:Collisions}, we directly state the equations of motion in terms of the four-momentum $k$ conjugate to $x-y$.
Similar to the 1PI calculation, we introduce the screened vertex
\begin{align}
\begin{split}
    \tilde{\Pi}^{R/A}(p)=\frac{1}{\frac{3N}{2\lambda}-\tilde{V}^{R/A}(p)}\\
    \tilde{\Pi}^K(p)=|\tilde{\Pi}^R(p)|^2 \tilde{V}^K(p)\,,
\end{split}
\end{align}
dressed by repeated transversal and longitudinal fluctuation bubbles
\begin{align}
\begin{split}
    \tilde{V}^{R/A}(k)=&i\left(G^{R/A}\star G^K+G^{A/R}\star G^V\right)(k)\\
    \tilde{V}^K(k)=&\frac{i}{2}\left( G^K\star G^K +G^V\star G^V\right.\\&\quad\left. +G^R\star G^R+G^A\star G^A\right)(k)\\
    \tilde{V}^V(k)=&i\left(G^K\star G^V+G^R\star G^A\right)(k)\,.
\end{split}
\end{align}
Here $(A\star B)(k)= \int_q A(k-q)B(q)$ indicates the convolution in frequency and momentum space. Furthermore, the tilde indicates that the projection to causal Green's functions has not been taken yet. We also note, that the diagrams contributing to $\tilde{\Pi}$ are the same that dress the vertex in random phase approximation and those found in leading order in $1/N$ for the 1PI transversal self-energy (see Eq.~\eqref{eq:1-PI_Vertex}). Here, however, the Green's functions within the fluctuation bubble are themselves dressed by fluctuation bubbles.
With this notation, and expanding the logarithm to linear order in $\tilde{V}^V$ one finds the simple expression
\begin{align}
    \Gamma_2^\text{RPA}=\!-\frac{i}{2}\!\int_k\!\left[\tilde{\Pi}^K(k)\tilde{V}^V(k)+\ln{\tilde{\Pi}^R(k)}+\ln{\tilde{\Pi}^A(k)}\right]\,,
\end{align}
which, however, is accurate only up to terms of order ${G^V}^2$ and $\left(G^R\star G^A\right)^2$, which do not contribute to the self-energy once evaluated for causal Green's functions.

\begin{figure}[ht]
\centering
\includegraphics[width=.9\columnwidth]{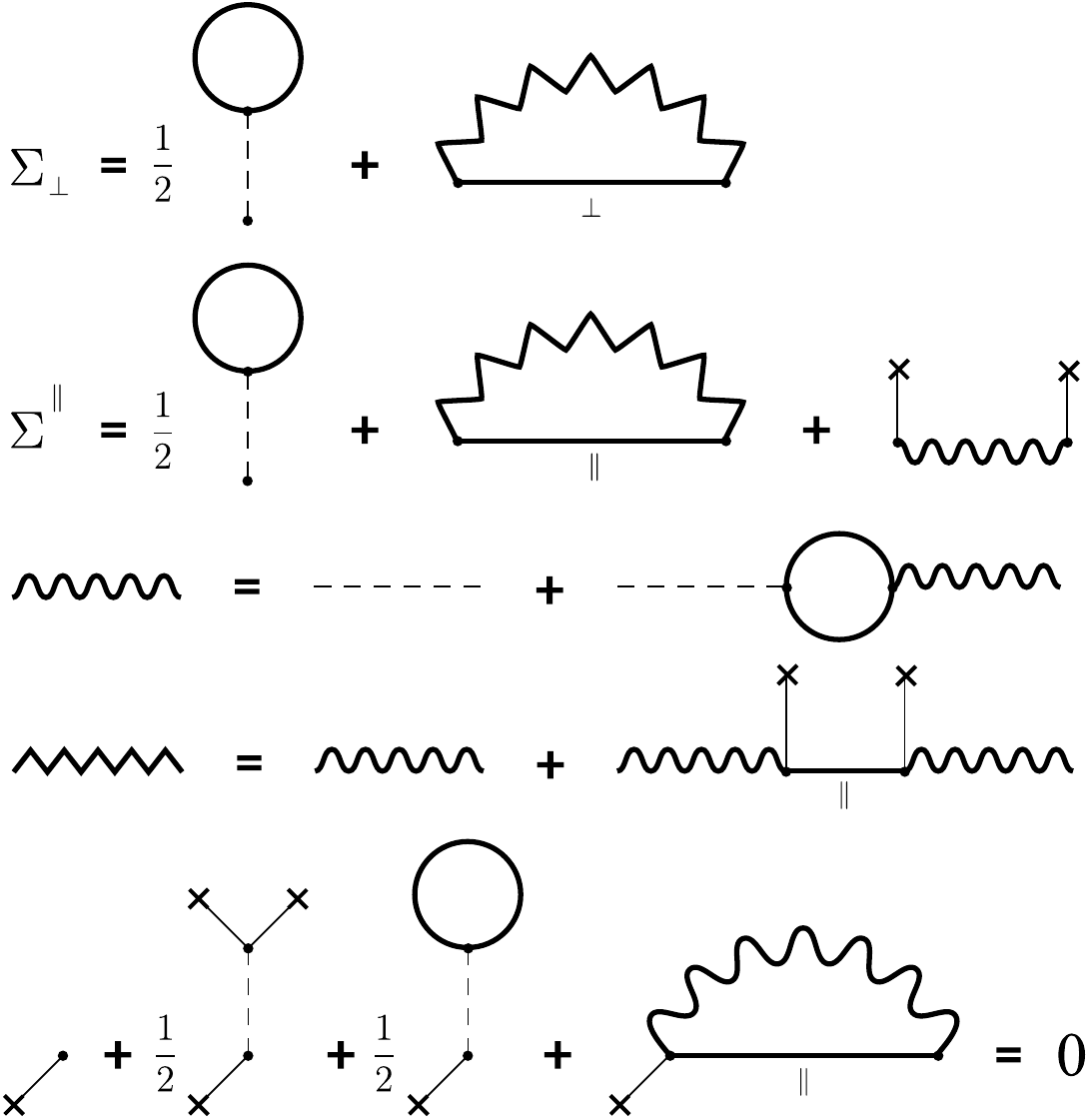}
\caption{Coupled Dyson equations in 2PI at next-to-leading order in $1/N$ with symmetry factors stated explicitly. To write the self-energies in a compact form we distinguish between two types of dressed interactions indicated as wavy and zigzag lines. The last line corresponds to the equation of motion of the field expectation value $\phi\equiv \phi_\text{cl}$.} 
\label{fig:Dyson}
\end{figure}

We again calculate the self-energies and obtain
\begin{align}\label{eq:2PI_selfenergy}
\begin{split}
%    \Sigma^R_\perp(k)=\,&\frac{i\lambda}{3N}\int_p G^K(p)+i \left(G^K_\perp\star\Pi^R+G^A_\perp\star\Pi^K\right)(k)+i\phi^2\left\{\left(\Pi G_\parallel \Pi\right)^R\star G^K_\perp+\left(\Pi G_\parallel \Pi\right)^K\star G^R_\perp\right\}(k)\\
%    \Sigma^R_\parallel(k)=\,&\frac{i\lambda}{3N}\int_p G^K(p)+i\left(G^K_\parallel\star\Pi^R+G^R_\parallel\star\Pi^K\right)(k)+\phi^2\left\{\Pi^R+i \left[\left(\Pi G_\parallel \Pi\right)^R\star G^K_\parallel+\left(\Pi G_\parallel \Pi\right)^K\star G^R_\parallel\right]\right\}(k)\\
%    \Sigma^K_\perp(k)=\,&i\left(G^R_\perp\star \Pi^R+G^A_\perp\star\Pi^A+G^K_\perp\star\Pi^K\right)(k)+i\phi^2\left\{\left(\Pi G_\parallel \Pi\right)^R\star G^R_\perp+\left(\Pi G_\parallel \Pi\right)^A\star G^A_\perp+\left(\Pi G_\parallel \Pi\right)^K\star G^K_\perp\right\}(k)\\
%    \Sigma^K_\parallel(k)=\,&i\left(G^A_\parallel\star\Pi^R+G^R_\parallel\star\Pi^A+G^K_\parallel\star\Pi^K\right)(k)\\&+\phi^2\left\{\Pi^K+i\left[\left(\Pi G_\parallel \Pi\right)^R\star G^R_\parallel+\left(\Pi G_\parallel \Pi\right)^A\star G^A_\parallel+\left(\Pi G_\parallel \Pi\right)^K\star G^K_\parallel\right]\right\}(k)\,,
    \Sigma^R_\perp(k)=\,&\frac{i\lambda}{3N}\int_p G^K(p)+i \left(G^K_\perp\star\Xi^R+G^A_\perp\star\Xi^K\right)(k)\\
    \Sigma^R_\parallel(k)=\,&\frac{i\lambda}{3N}\int_p G^K(p)+i\left(G^K_\parallel\star\Xi^R+G^R_\parallel\star\Xi^K\right)(k)\\&+\phi^2\Pi^R(k)\\
    \Sigma^K_\perp(k)=\,&i\left(G^R_\perp\star \Xi^R+G^A_\perp\star\Xi^A+G^K_\perp\star\Xi^K\right)(k)\\
    \Sigma^K_\parallel(k)=\,&i\left(G^A_\parallel\star\Xi^R+G^R_\parallel\star\Xi^A+G^K_\parallel\star\Xi^K\right)(k)\\&+\phi^2\Pi^K(k)\,,
\end{split}
\end{align}
where
\begin{align}\label{eq:Xi}
\begin{split}
    \Xi^{R/A}(k)&=\Pi^{R/A}(k)\left(1+\phi^2\Pi^{R/A}(k)G^{R/A}_\parallel(k)\right)\\
    \Xi^K(k)&=\Pi^K(k)+\phi^2\left(\Pi G_\parallel \Pi\right)^K(k)
\end{split}
\end{align}
represents the interactions screened by both transverse and longitudinal fluctuations. Here we have abbreviated the Keldysh structure with $(A B C)^K=A^R B^R C^K+A^R B^K C^A+A^K B^A C^A$. Also, $\Pi$ has been projected to the physical state satisfying the causality condition $G^V_{\perp/\parallel}=0$ and
\begin{align}
\begin{split}
    G^{R/A}_{\perp/\parallel}(k)&=\left(\left[G_{0,\parallel}^{R/A}\right]^{-1}-\Sigma^{R/A}_{\perp/\parallel}(k)\right)^{-1}\\
    G^K_{\perp/\parallel}(k)&=|G^R_{\perp/\parallel}(k)|^2\left(\Sigma^K_{\perp/\parallel}(k)-8i\gamma T\right)\,.
\end{split}
\end{align}
In the absence of fluctuations the inverse transversal Green's function is given by
\begin{align}
    \left[G_{0,\perp}^{R/A}\right]^{-1}(k)=2\left(\omega^2+i\gamma\omega-m-D \vect{k}^2\right)-\frac{\lambda}{3N}\phi^2
\end{align}
and the difference between longitudinal and transversal modes has been absorbed into the self-energy.
Additionally, the equation for the field expectation value becomes
\begin{align}\label{eq:2PI_phi}
\begin{split}
    &\left(\left[G_{0,\perp}^{R/A}\right]^{-1}(k)-\frac{i\lambda}{3N}\int_q G^K(q)\right.\\&\qquad\quad\left.-i \left(G^K_\parallel\star\Pi^A+G^R_\parallel\star\Pi^K\right)(k)\right) \phi(k)=0\,.
\end{split}
\end{align}
While these equations are lengthy, this is owed mostly to the explicit distinction between longitudinal and transversal modes. In fact, there is a simple diagrammatic representation for these coupled Dyson equations shown in Fig.~\ref{fig:Dyson}.

\begin{figure}[ht]
\centering
\includegraphics[width=.3\columnwidth]{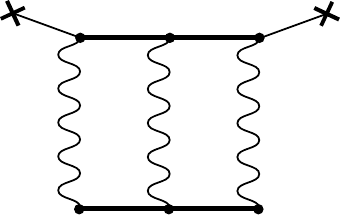}
\caption{Contribution missing from the transversal self-energy responsible for the leading contribution to the Goldstone mass of order $\mathcal{O}(N^{-2})$.} 
\label{fig:missing}
\end{figure}

\begin{figure}[ht]
\centering
\includegraphics[width=.5\columnwidth]{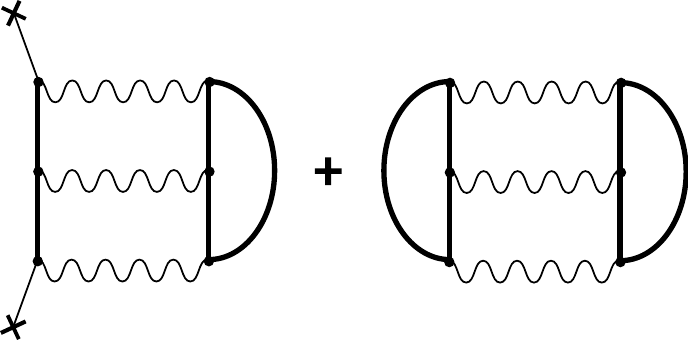}
\caption{Feynman diagrams that need to be added to the effective action to suppress the Goldstone mass in 2PI to $\mathcal{O}(N^{-3})$.} 
\label{fig:extension}
\end{figure}

From there we immediately see that Goldstone's theorem reduces to equating the last diagram in the first and last lines at vanishing external momentum. However, as opposed to 1PI, self-energies calculated from a 2PI effective action mix between different powers of $1/N$, therefore it is not at all obvious that these diagrams are identical. Indeed, one may compare the explicit expansions of both expressions order by order in $1/N$ and find that they disagree at $\mathcal{O}(N^{-2})$ because the diagram in Fig.~\ref{fig:missing} is missing from the transversal self-energy. The Goldstone mass could therefore be suppressed to $\mathcal{O}(N^{-3})$ by adding the diagrams in Fig.~\ref{fig:extension} to the effective action. By repeating this procedure, the violation of Goldstone's theorem can be suppressed to higher and higher orders in $1/N$ at the cost of calculating more and more increasingly complicated diagrams.

The issue of a finite gap in the symmetry-broken phase for conserving approximations has been known for many years \cite{Baym1977, Haussmann2007}, but no solution is in sight. Fortunately, the erroneous mass is small $\mathcal{O}(N^{-2})$. We will thus in the following use the 2PI effective action and manually set the transverse mass in thermal equilibrium to zero as this is much more straightforward than attempting to estimate and compensate the errors of a non-conserving approximation.

Until now the discussion has been completely generic and equally applies to equilibrium, however (approximate) conservation laws due to algebraically slow decay and massless modes will remain the central concepts for subsequent approximations needed to evolve the system to late times.

\section{Wigner expansion}\label{sec:Wigner}

Within any given approximation to $\Gamma_2$, the evolution of the system is given by the Dyson equation~\eqref{eq:Dyson}, which can be rewritten as a set of coupled integro-differential equations. These are however difficult to interpret and in general numerically very costly to evolve to late times. Therefore it necessitates further approximations \cite{Lipavsky1986, Aoki2014, Joost2020, Kaye2021}. A very useful tool in this context is the Wigner expansion, a gradient expansion on the Wigner transformed Green's functions
\begin{align}
    G(t,\omega,\vect{k})=\int_{-\infty}^\infty d\tau e^{i\omega \tau}G\left(t+\frac{\tau}{2},t-\frac{\tau}{2},\vect{k}\right)\,.
\end{align}
As is usually the case, we expand to linear order in $\partial_t$. Furthermore, spatial translation invariance is assumed at all times.
A brief summary of the Wigner expansion is provided in App.~\ref{app:Wigner}. Within the Wigner approximation the retarded Green's function simply becomes
\begin{align}\label{eq:t-evoGR}
    G^R(t,\omega,\vect{k})=\left(\left[G^R_0\right]^{-1}(\omega,\vect{k})-\Sigma^R(t,\omega,\vect{k})\right)^{-1}.
\end{align}
The Keldysh Green's function satisfies the time-local open Kadanoff-Baym equation derived below in Sec.~\ref{sec:SingleMode}
\onecolumngrid
\begin{align}\label{eq:KB_Wigner}
\begin{split}
    \gamma\partial_t G^K=&-\left(2(D \vect{k}^2+m)+\frac{\lambda}{3N}\phi^2_\text{cl}+\Re{\Sigma^R}\right)G^K-\Re{G^R}\left(\Sigma^K-8i\gamma T\right)\\&-\frac{1}{2}\{\Im{\Sigma^R},G^K\}+\frac{1}{2}\{\Im{G^R},\Sigma^K\}+\Sigma^R_\text{mem} G^K-\Re{G^R}\Sigma^K_\text{mem}
\end{split}
\end{align}
\twocolumngrid
\noindent with the Poisson bracket $\{A,B\}=\partial_\omega A\partial_t B-\partial_t A \partial_\omega B$ and $\Sigma^{R/K}_\text{mem}$ representing corrections to the purely time local interaction terms of the first line. They are obtained from a Taylor expansion of the dynamic functions within the integral expressions for $\Sigma^{R/K}$ and therefore correspond to memory effects \cite{Knoll2001}. The evolution is conserving if the expressions for the
self-energies are derived from a 2PI effective action and can be made gapless by
subtracting the equilibrium gap $\Sigma^R(t\to\infty,\omega=0,\vect{k}=0)$. Note, that this only affects the time-local terms in the first line of Eq.~\eqref{eq:KB_Wigner}.

Scaling dynamics implies algebraic relaxation $\sim t^{-\alpha}$, which suggests that gradient terms are suppressed at late times. We will analyze this argument more carefully in Sec.~\ref{sec:Gapless} but for the moment will push on by dropping all gradient terms involving the self-energy. With this, we find the evolution equation
\begin{align}\label{eq:t-evo}
    Z^{-1}\partial_t G^K=&2i\Re{\left({G^R}^{-1}\right)}G^K-2i\Re{G^R}\left(\Sigma^K-8i\gamma T\right)
\end{align}
with inverse quasiparticle weight $Z^{-1}=2i\gamma$. It has a simple interpretation, with the first term corresponding to the insertion of fluctuations at momentum $\mathbf{k}$ either from the bath or by internal collisions leading to the redistribution of excitations from other momenta. The second term on the other hand describes the opposite effect of excitations decaying into the bath or scattering to other momenta.
We note that this equation differs qualitatively from its closed system equivalent
\begin{align}\label{eq:t-evo_closed}
    \partial_t G^K=-2\left(\Im{\left({G^R}^{-1}\right)}G^K+\Im{G^R}\Sigma^K\right)\,,
\end{align}
where the real part of $G^R$ is replaced with the imaginary part. This reflects the qualitative difference in the late-time evolution we already pointed out in Sec.~\ref{sec:Intuition}. The fact that the Wigner expansion is applicable only at late times rationalizes that the transition from closed-system to open-system dynamics at time scales $\sim 1/\gamma$ are beyond reach, leading to the observed dichotomy between the two settings (see Sec.~\ref{sec:SingleMode} below). 

The key feature of the Wigner expansion is the simplification of the memory integrals to purely time-local terms in Wigner coordinates. We, therefore, refer to Eq.~\ref{eq:t-evo} as memoryless evolution.

Before the Wigner expansion can be applied, we need to address a few issues: firstly, we need to verify if it is actually applicable at late times for gapless approximations, secondly, we need to confirm that it is conserving and does not introduce an artificial gap to the transverse modes. We address the first issue in Sec.~\ref{sec:Gapless}, arguing that it produces the correct qualitative behavior at late times. I.e., scaling exponents will be captured correctly if the time-dependent momentum scale that dominates the collisions decays as $k_\text{Coll}\gtrsim t^{-1/z}$ with $z$ the dynamical exponent. Observables local in frequency space on the other hand are not expected to be accurate unless $k_\text{Coll}\gg t^{-1/z}$. 

The second issue has been addressed by Knoll et al. \cite{Knoll2001}, who have shown the gradient expansion to be conserving at every order. Intuitively this is easy to understand, as for any frequency and momentum we can consider sufficiently late times, such that the evolution has slowed down enough such that a scale separation between different orders of time derivatives occurs \footnote{Strictly speaking this simple argument fails since the Wigner expansion is an asymptotic series, necessitating the more involved discussion of Knoll et al. \cite{Knoll2001}.}. Since the cancellation happens independently for each diagram in the 2PI effective action, the order of the expansion can be chosen differently for each scattering process. In particular, one may keep gradient corrections to the bare dynamics but only treat the lowest order of the collision integral as is done in Eq.~\eqref{eq:t-evo}. Following the same arguments, it also follows that the Wigner expansion does not introduce an artificial gap. 

As a side note, we point out, that the truncation of the collision integral at zeroth order in the Wigner expansion necessarily fixes the inverse quasiparticle weight, i.e. the prefactor of $\partial_t G^K$ in Eq.~\eqref{eq:t-evo} to its bare value $Z^{-1}=2i\gamma$. Attempting to improve this value to $Z^{-1}=2i\gamma+i\partial_\omega \Im{\Sigma^R}$ as is often done \cite{Kamenev_book, Wasak2022} will break conservation laws unless it is the same for all quasiparticles. Vice versa, if the quasiparticle weight differs significantly from its bare value, memory effects and backflow terms in the form of the second line in Eq.~\eqref{eq:KB_Wigner} must be kept. For the qualitative analysis that we will get to next, the precise value of $Z$ is insignificant.

We conclude that the quench dynamics of an open system at late times in the symmetry-broken phase is adequately described by the evolution equations Eqs.~\eqref{eq:t-evoGR}\eqref{eq:t-evo} if self-energies are derived from a 2PI effective action, the inverse quasiparticle weight in equilibrium is close to its bare value $Z^{-1}\approx 2i\gamma$ and the characteristic scale of the momentum distribution of fluctuations $k_\text{Coll}(t)$ does not relax more quickly than $t^{-1/z}$. In the case discussed here, the dynamical exponent takes the value $z=2$.

\subsection{Overdamped dynamics}\label{sec:SingleMode}

The necessity for an explicit distinction between open and closed evolution equations in the memoryless approximation becomes apparent already for a non-interacting system, as will be explained in this subsection. 

For each Green's function $G(t_1,t_2)$, there are two Dyson equations: one for the time evolution of each of its arguments. The sum and difference of these provide two linearly independent evolution equations. For the Keldysh Green's function, these are known as Kadanoff-Baym equations ($\circ$ denotes the convolution in time)
\begin{align}\label{eq:KB}
\begin{split}
 [G^R]^{-1}\!\circ G^K\pm G^K \!\circ [G^A]^{-1}=&\Sigma^K\!\circ G^A\pm G^R\!\circ\Sigma^K\\&-8i\gamma T \left(G^A\pm G^R\right)\,,
\end{split}
\end{align}
which follow directly from the Dyson equation \eqref{eq:Dyson} and are therefore exact.

It is often convenient to distinguish between terms of the self-energies that are local in time and nonlocal ones. The former, so-called Hartree terms $\Sigma_H^{R/A}(t,t')\sim \delta(t-t')$ are responsible for a dynamical modification of the spectrum. The time nonlocal terms $\Sigma_\text{nl}^{R/A}=\Sigma^{R/A}-\Sigma_H^{R/A}$ on the other hand describe the effect of collisions and cause redistribution in momentum space. They are therefore the sole reason for thermalization in closed systems at late times after a quench. These two very different effects of interactions are often made explicit in the Kadanoff-Baym equations by collecting the non-local terms in the so-called collision integral $I$. One then has
\begin{align}
\begin{split}
 ([G_0^R]^{-1}-\Sigma_H^R)\!\circ G^K\pm G^K \!\circ ([G_0^A]^{-1}-\Sigma_H^A)\\=I-8i\gamma T \left(G^A\pm G^R\right)\,,
\end{split}
\end{align}
with the collision integral
\begin{align}\label{eq:Icoll}
    I=\Sigma_\text{nl}^R\circ G^K\pm G^K\circ \Sigma_\text{nl}^A+\Sigma^K\circ G^A\pm G^R\circ \Sigma^K\,.
\end{align}

The two signs in Eq.~\eqref{eq:KB} lead to different dynamics in Wigner approximation \footnote{For complex fields only the minus sign in \eqref{eq:KB} leads to a 
differential equation in time and is therefore exclusively considered.}.
To understand which is more suitable to us, we consider a damped harmonic oscillator 
(i.e. the open $O(1)$-model with $\lambda=0$), for which the collision integral vanishes. It is described by the two equations
\onecolumngrid
\begin{align}
    \left(\partial_t+\gamma\right)G^K(t,\omega,\vect{k})&=-\frac{2i\gamma^2 T}{\left(\omega^2-D \vect{k}^2-m\right)^2+\gamma^2\omega^2}\label{eq:KB+}\\
    \left[\partial_t^2+2\gamma\partial_t-4\left(\omega^2-D \vect{k}^2-m\right)\right]G^K(t,\omega,\vect{k})&=\frac{8i\gamma T\left(\omega^2-D \vect{k}^2-m\right)}{\left(\omega^2-D \vect{k}^2-m\right)^2+\gamma^2\omega^2}\label{eq:KB-}\,.
\end{align}
\twocolumngrid
Surprisingly \eqref{eq:KB-} is unstable for $\omega^2>m+D\vect{k}^2$, a property that is
clearly not shared by the damped harmonic oscillator, yet \eqref{eq:KB-} is exact. In the exact evolution, this issue is resolved by the initial conditions, which guarantee zero overlap with the unstable
eigenfunctions. However, the Wigner representation is not well suited to implement these boundary conditions, which there take the form of integral equations that are not easily generalized to the interacting system. 
We may avoid these complications by considering the overdamped limit, where \eqref{eq:KB-}
simplifies to
\begin{align}\label{eq:Overdamped}
    \left[\partial_t+\frac{2}{\gamma}\!\left(D \vect{k}^2+m\right)\!\right]\!G^K(t,\omega,\vect{k})&=-\frac{4i T\left(D \vect{k}^2+m\right)}{\left(D \vect{k}^2+m\right)^{\!2}\!+\gamma^2\omega^2}\,.
\end{align}
Physically, this neglects the quickly decaying mode with lifetime $\sim 1/\gamma$, leaving only the long-lived branch with decay rate $\sim \vect{k}^2$, and is thus accurate on time scales $t \gg 1/\gamma$.
For the closed system, however, no such limit exists as there $\gamma=0^+$. Consequently, the undamped harmonic oscillator needs to be evolved with \eqref{eq:KB+} in the limit $\gamma\to 0^+$, which only becomes non-trivial once interactions are included. Applying this conclusion to the evolution of closed systems in general leads to Eq.~\eqref{eq:t-evo_closed}.

For open systems, on the other hand, we have the choice between Eqs.~\eqref{eq:KB+} and \eqref{eq:Overdamped}. To understand, which is more suitable for systems with gapless modes, we generalize to the driven case with time-dependent mass and temperature and integrate over all frequencies. As long as the frequency integral is convergent, this requires no approximation as all gradient terms from the Wigner expansion can be written as total derivatives.
In case of Eq.~\eqref{eq:KB+} this is only true for a positive mass $m(t)>0$ and the integrated equation
\begin{align}
    \left(\partial_t+\gamma\right)G^K(t,\vect{k})=-i\frac{\gamma T(t)}{D \vect{k}^2+m(t)}
\end{align}
diverges and becomes meaningless for $m(t)<0$. The integrated, overdamped
limit of \eqref{eq:KB-} on the other hand becomes
\begin{align}\label{eq:Overdamped_int}
    \left[\partial_t+\frac{2}{\gamma}\!\left(D \vect{k}^2+m\right)\!\right]\!G^K(t,\vect{k})&=-\frac{2iT}{\gamma}\,,
\end{align}
which clearly has no such issues. It is stable for all times and drives and becomes exact for quenches in the limit $t\to\infty$. Indeed, Eq.~\eqref{eq:Overdamped_int} corresponds to the time-dependent stochastic Ginzburg-Landau equation in the limit $\lambda=0$, which we show in Sec.~\ref{sec:MF}
to be very accurate at late times, even for strong interactions. We therefore conclude, that at first order in the Wigner expansion, gapless, open systems have to be evolved with the opposite sign choice in the Kadanoff-Baym equation \eqref{eq:KB} than closed systems.

The fact that the dynamics of the open, gapless
system is faithfully captured only by \eqref{eq:KB-} introduces the previously stated dichotomy 
between the kinetic theories for closed and open systems. The kinetic theory is therefore
in general ill-suited to discuss a crossover between the dynamics of closed and open $O(N)$ models.

Clearly, this argument remains valid when adding interactions. However, care has to be taken when applying the Wigner expansion to the right-hand side of Eq.~\eqref{eq:KB} as the gradient expansion needs to be applied to the loop integrals in the self-energies as well (see \cite{Knoll2001}). With the systematic application of the Wigner expansion also to those terms, the Kadanoff-Baym equation for open $O(N)$-symmetric systems takes the general form stated in Eq.~\ref{eq:KB_Wigner}.

We conclude the paragraph with a technical remark. As we argued, the overdamped approximation is accurate at small momenta. The frequency integrals of loop diagrams however are unrestricted. It is, therefore, necessary to first use exact cancellations due to the causality structure of the Keldysh Green's functions such as $\int_\omega (G^R(\omega,\vect{k})+G^A(\omega,\vect{k}))=0$ before the overdamped approximation is applied. Following these cancellations only integrals regularized by the distribution function remain and since the quickly decaying gapped mode is essentially unoccupied at sufficiently late times, the approximation can be safely applied there.

\subsection{Memoryless collisions}\label{sec:Gapless}
To reach beyond the non-interacting, Markovian evolution discussed in the previous subsection, it is necessary to apply the Wigner expansion also to the collision integral $I$ defined in Eq~\eqref{eq:Icoll}. As we have argued before, one naively expects for self-similar dynamics that the derivative terms in the second line of \eqref{eq:KB_Wigner} disappear faster than the interaction terms in the first line. However, we are going to find that the relaxational dynamics towards smaller momenta also evolves the time scale beyond which this argument holds. For a Gaussian, thermal fixed point, meaning that collisions are irrelevant in the approach to the stationary state, the occupation induced by a quench vanishes as $|\vect{k}|\sim t^{-1/z}$, which coincides with the limit of validity of the Wigner expansion. In this marginal case, realized for the open $O(N)$ model, the collision integral is captured qualitatively by the memoryless approximation if either the relaxation is slower than $1/t$ or the interactions are sufficiently strongly screened. In the first case, the evolution becomes adiabatic, while in the second case, the memory is cut off by a time scale $\xi^z$ determined by the screening length $\xi$.

We generally expect all terms in the second line of Eq.~\eqref{eq:KB_Wigner} to scale similarly but note that a discussion of all terms requires their explicit evaluation, which then becomes specific for the choice of model and approximation to the 2PI effective action. On the other hand, we provide sufficient, but not necessary conditions for convergence of the Wigner expansion. It is therefore plausible that memoryless approximations are in fact more broadly applicable. 

For concreteness, we will only consider the Wigner expansion of the driving term of the collision integral i.e. the last term of Eq.~\eqref{eq:t-evo}. Furthermore, we focus on the special, relevant case, where $G^R(t,\omega,\vect{k})=G^R(\omega,\mathbf{k})$ at late times is given by Eq.~\eqref{eq:GRqp} with $\gamma=D=1$ and $\mu=0$. In NLO($1/N$) it is then expected that the other terms behave similarly. Moreover, we assume that $\Sigma^K(t,\omega,\vect{k})$ varies in frequency on a scale $\omega_0$, which means that we need to distinguish between high and low momenta.

\begin{figure}[ht]
    \centering
    \includegraphics[width=\columnwidth]{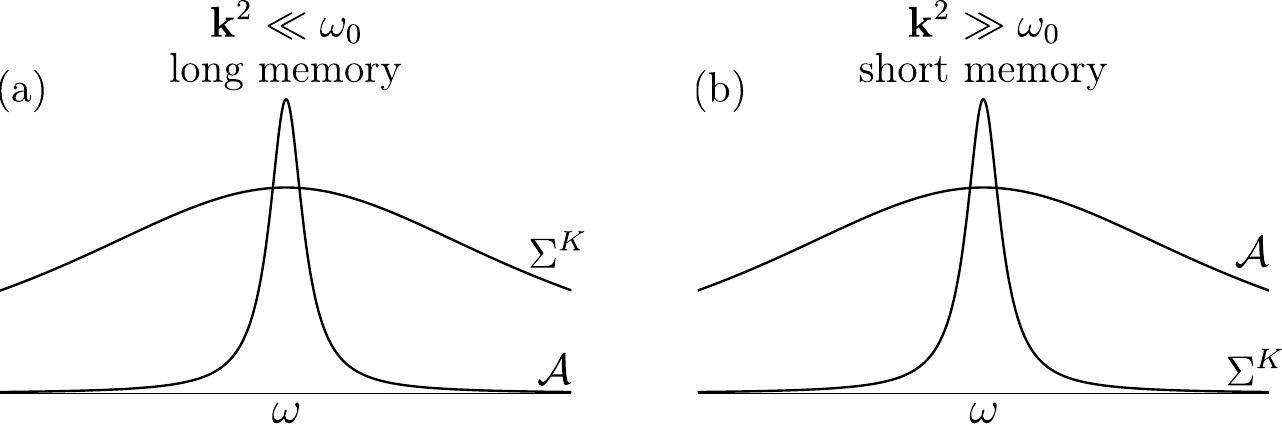}
    \caption{Comparison between the two limiting cases of small and large momenta. At small momenta (a), the correlations of $\Sigma^K(t_1,t_2,\mathbf{k})$ are very short compared to those in the spectral function $\mathcal{A}=G^A-G^R$. Hence the dependence of $\Sigma^K(t_1,t_2,\mathbf{k})$ on $t_1-t_2$ can be approximated as $\delta(t_1-t_2)$, which in frequency space translates to $\Sigma^K(t,\omega,\vect{k})=\Sigma^K(t,\mathbf{k})$. However, its algebraic dependence on $t$ remains important as the narrow spectrum implies a long memory. In the opposite case (b) excitations are very short-lived, therefore the slow variation of $\Sigma^K(t,\omega,\vect{k})$ in time can be neglected, but its relatively long correlations in $t_1-t_2$ render its frequency dependence in Wigner coordinates important. The two cases therefore can be thought of as a piano with a pressed damper pedal (a) or una corda pedal (b).}
    \label{fig:GKBAvsWigner}
\end{figure}

\emph{High momenta ---} In the regime $|\mathbf{k}|^z\gg \omega_0$ excitations are short-lived, such that we can approximate $\Sigma^K$ in the collision integral by its instantaneous value
\begin{align}
I(t,\omega,\vect{k})\approx G^R(\omega,\mathbf{k})\Sigma^K(t,\omega,\vect{k})\quad \text{for}\quad t\gg1/\omega_0\,,    
\end{align}
which implies that the lowest order Wigner expansion becomes accurate after a fixed finite time, see Fig.~\ref{fig:GKBAvsWigner}. We note, that the frequency dependence of $\Sigma^K$ is essential at high momenta. Therefore, the generalized Kadanoff-Baym ansatz, which approximates the Keldysh Green's function by the covariance matrix (i.e. its equal time limit) \cite{Lipavsky1986} overestimates the population at high momenta.

\emph{Low momenta ---} For $|\mathbf{k}|^z\ll \omega_0$ the Keldysh self-energy can be approximated as a constant in frequency
\begin{align}\label{eq:ToySigmaK}
\Sigma^K(t,\omega,\vect{k})\approx\Sigma^K(t,\mathbf{k})\underset{t\to\infty}{\longrightarrow} t^{-\alpha}    
\end{align}
that decays algebraically at late times with $\alpha >0$. This case implies that the driving saturates the linewidth of the spectral function which describes long-lived excitations that remember the history of $\Sigma^K$, see Fig.~\ref{fig:GKBAvsWigner}. This corresponds to the generalized Kadanoff Baym ansatz. Unfortunately, there is no systematic way to interpolate between the generalized Kadanoff-Baym ansatz at small momenta $|\mathbf{k}|^z\ll \omega_0$ and the Wigner expansion at high momenta $|\mathbf{k}|^z\gg \omega_0$. Instead, we are going to derive the conditions under which the Wigner expansion can be applied at small momenta.

We first note that for the exemplary contribution to the collision integral the Wigner expansion to $n$-th order can be solved exactly for $|\mathbf{k}|^z\ll \omega_0$, where $\Sigma^K$ becomes frequency independent. There one finds
\onecolumngrid
\begin{align}
\begin{split}
I_n(t,\omega,\vect{k})&=\sum_{m=0}^n\frac{i^m}{2^m m!}\partial^m_\omega\left(G^R(\omega,\mathbf{k})+(-1)^m G^A(\omega,\mathbf{k})\right)\partial^m_t\Sigma^K(t,\mathbf{k})\\&=-\sum_{m=0}^n t^{-m-\alpha}\left((|\mathbf{k}|^z-i\omega)^{1+m}+(|\mathbf{k}|^z+i\omega)^{1+m}\right)(2(|\mathbf{k}|^{2z}+\omega^2))^{-1-m}(\alpha)_m\,,
\end{split}
\end{align}
where $(\alpha)_n=\Gamma(\alpha+n)/\Gamma(\alpha)$ is  the Pochhammer symbol. This series may be compared with the exact result (for $\Sigma^K(t,\mathbf{k})=t^{-\alpha}$)
\begin{align}
\begin{split}
    I(t,\omega,\vect{k})&\equiv\int d\tau e^{i\omega \tau}G^R(\tau,\mathbf{k})\Sigma(t-\tau/2,\mathbf{k})\\&=e^{-2t(|\mathbf{k}|^z+i\omega)}t^{1-\alpha}\left(e^{4it\omega}E_\alpha(-2t(|\mathbf{k}|^z-i\omega))+E_\alpha(-2t(|\mathbf{k}|^z+i\omega))\right)\,.
\end{split}
\end{align}
The first observation is that the Wigner expansion for any fixed set of arguments is an asymptotic series that for large $n$ diverges as $$|I_n(t,\omega,\vect{k})|=n^n (2 e t)^{-n} (|\mathbf{k}|^{2z} + \omega^2)^{-n/2}$$ to logarithmic accuracy. More importantly, the difference between the Wigner expansion and the exact result is largest for $\omega=0$, where it behaves as
\begin{align}
    \left|\frac{I(\omega=0,t,\mathbf{k})-I_n(\omega=0,t,\mathbf{k})}{I(\omega=0,t,\mathbf{k})}\right|\sim\begin{cases}
        (t|\mathbf{k}|^{z})^{-\alpha-n}&t|\mathbf{k}|^z\ll 1\\
        (t|\mathbf{k}|^{z})^{-1-n}&t|\mathbf{k}|^z\gg 1
    \end{cases}
\end{align}
\twocolumngrid
%is small only if $t\gg \min\{|\mathbf{k}|^{-\frac{2n+2}{n+\alpha}},|\mathbf{k}|^{-\frac{2n+4}{n+1+\alpha}}\}\underset{n\to 0}{\longrightarrow}\min\{|\mathbf{k}|^{-2/\alpha},|\mathbf{k}|^{-4/(1+\alpha)}\}$.
\noindent and is therefore small only if $t\gg |\mathbf{k}|^{-z}$.
The behavior for large $n$ is easy to understand: The Green's function provides a memory time $t_\text{mem}\sim |\mathbf{k}|^{-z}$. If $t\lesssim t_\text{mem}$ the collision integral remembers the singularity at $t=0$ and the gradient expansion will never converge. In fact, each successive order will deviate further from the actual result. 

So far, we have found that the zeroth order Wigner expansion is accurate at high momenta $|\vect{k}|^z\gg \omega_0$ and for small momenta $|\vect{k}|^z\ll \omega_0$ if $|\vect{k}|\gg t^{-1/z}$. However, this excludes the particularly important case of canonical scaling $|\vect{k}|\sim t^{-1/z}$, for which we will explain next that the Wigner expansion remains valid for quantities averaged over a sufficiently large frequency range.

\begin{figure}[ht]
    \centering
    \includegraphics[width=\columnwidth]{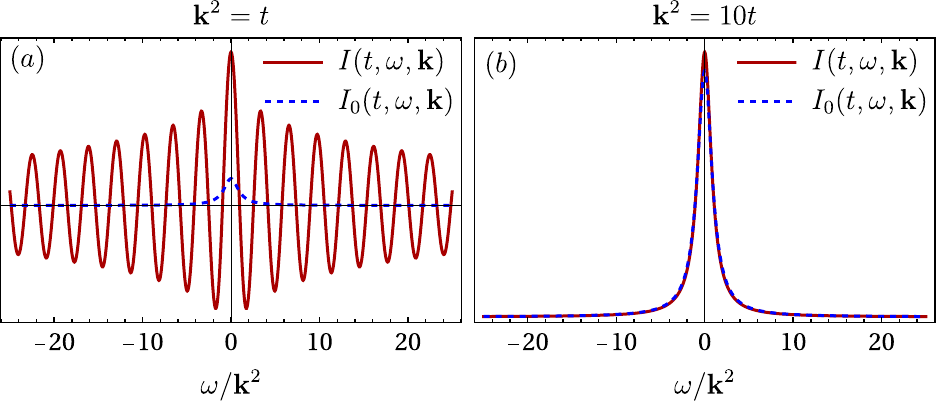}
    \caption{Comparison between the full integral $I(t,\omega,\vect{k})$ for $t=100$, $\alpha=-3/2$ and $z=2$ with the zeroth order Wigner approximation $I_0(t,\omega,\vect{k})$. In (a) we chose the momentum $|\mathbf{k}|=\sqrt{t}$, where the Wigner expansion is locally not yet a good approximation as it misses the fast oscillations $\sim \cos{(2 \omega t)}$. At the same time at the larger momentum $|\mathbf{k}|=\sqrt{10 t}$ shown in (b) the system has already forgotten about the quench and is therefore well approximated by the zeroth order Wigner expansion at all frequencies. At small momenta one thus has to average over a frequency range $\gg 1/t$ for the Wigner expansion to become a good approximation.}
    \label{fig:Wigner_Test}
\end{figure}

When discussing times $t\sim t_\text{mem}$ we need to extend the discussion to include the initial quench. To illustrate this, we consider $\Sigma^K(t)=t^{-\alpha}e^{-\delta t/t}\theta(t)$, which is a smooth (non-analytic) function that is exactly zero at negative times and then switches on on a time scale $\delta t=1$. Since $\Sigma^K$ is independent of frequency, it corresponds to Markovian noise. At times $t\gg t_\text{mem}$ the zeroth order Wigner expansion is well-converged (see Fig.~\ref{fig:Wigner_Test}). But for earlier times the system remembers the quench process, which results in oscillations of the form $I(t,\omega,\vect{k})\sim \cos{(2\omega (t+\mathcal{O}(\delta t)))}$ (see Fig.~\ref{fig:Wigner_Test}). While locally the Wigner expansion is almost everywhere a bad approximation at these early times/small momenta, we notice that the frequency integrated term
\begin{align}
\begin{split}
    I(t,\mathbf{k})&=\int\frac{d\omega}{2\pi}I(t,\omega,\vect{k})\\&=\left(G^R(t=0,\mathbf{k})+G^A(t=0,\mathbf{k})\right)\Sigma^K(t,\mathbf{k})\\&=-\frac{1}{2}\Sigma^K(t,\mathbf{k})\equiv I_0(t,\mathbf{k})
\end{split}
\end{align}
is recovered exactly in zeroth order Wigner expansion (all other expansion terms can be written as integrals over total derivatives and thus vanish). Hence we conclude that frequency-integrated quantities are well described in zeroth order Wigner expansion. However internally in the collision integrals the frequency-dependent, oscillating Green's functions enter. Since the latter inherit these oscillations from the memory effects of the collision integral, we have to consider integrals of the form
\begin{align}\label{eq:collInt}
    I(t,\vect{k})=\int\frac{d\omega}{2\pi}I(t,\omega,\vect{k})K(\omega)\,.
\end{align}
Linearizing the collision integral around the thermal state, the kernel $K(\omega)$ is unaffected by memory effects and generally independent of time.

We now have to distinguish several cases:
\begin{itemize}
    \item Assuming canonical scaling, $K(\omega)$ varies on a scale $\sim k^{z}$ and since the relevant momenta depend on time one has $k^{z}\sim 1/t$. In this case, the zeroth order Wigner expansion captures the integral accurately only if $\alpha<1$, which corresponds to adiabatic evolution. For $\alpha=1$ memory effects give rise to logarithmic terms in Eq.~\eqref{eq:collInt}, which are missed by the Wigner expansion. In case of a fast relaxation $\alpha > 1$, the Wigner expansion predicts qualitatively wrong behavior.
    \item If $K(\omega)$ varies on a scale much smaller than $k^{z}$, the effect of the momentum integration is negligible and the Wigner expansion fails for $t\sim t_\text{mem}$
    \item Finally, if the characteristic scale of $K(\omega)$ is independent of time, the zeroth order Wigner expansion of Eq.~\eqref{eq:collInt} is accurate for all values of $\alpha$. In the case of the $O(N)$ model, this frequency scale is provided by the screening length $\xi^{-z}$. Since $\alpha=3/2$, screening is essential for the application of the formalism to the symmetry-broken phase of the $O(N)$ model.
\end{itemize}
%Not quantitative because of seed from early dynamics and validity of Wigner expansion only under the integral.

%These integrals are therefore only approximated well in Wigner expansion if they contain a factor that varies slowly on the scale of the quench-induced oscillations, i.e. $\delta \omega\gg 1/t$. With the dynamical exponent $z$ this translates to momenta $\mathbf{k}\gg t^{-1/z}$. Consequently, if the dominant contribution to collision integrals comes from momenta $\mathbf{k}\gg t^{-1/z}$, the Wigner expansion is well justified, and all quantities with a frequency resolution less than $1/t$ can be predicted accurately.

In summary, the applicability of the Wigner expansion to gapless open systems requires a case-by-case analysis, weighing the memory time of the most relevant momenta at late times against the elapsed time. For $t\sim t_\text{mem}$, which is realized for canonical scaling with $|\vect{k}|\sim t^{-1/z}$, one finds the very interesting case of a marginal expansion that recovers the correct scaling behavior only for frequency averaged quantities. This is sufficient for the scaling analysis of the late-time evolution and the purposes of this paper, as in the near-thermal, symmetry-broken state of the $O(N)$ model collisions are sufficiently screened at long distances. As is illustrated in Fig.~\ref{fig:Wigner_Test}, observables that are not averaged over a frequency range greater than $1/t$ retain a memory of the initial quench. These cannot be resolved in the Wigner approximation.

We note, that the discussion here is no proof of the applicability of the Wigner expansion but provides useful insights into the convergence properties of the collision integral. The simple arguments presented here do not account for cancellations between different contributions to the collision integral. We therefore consider the estimates presented here to be rather cautious and justify the use of the Wigner expansion accordingly. This applies especially to the collision integral in the approach to a Gaussian fixed point and thus in the open $O(N)$ model in the considered physical situation near thermal equilibrium. Consequently, in Eq.~\eqref{eq:t-evo} we neglect the second line of Eq.~\eqref{eq:KB_Wigner}, which in general has to be justified \emph{a posteriori}. This is done by validating that the relevant momenta of the collision integrals satisfy $k_\text{Coll}\gtrsim t^{-1/z}$ with special care needed in the marginal case $k_\text{Coll}\sim t^{-1/z}$. For a given solution, $k_\text{Coll}$ can be estimated by maximizing the functional derivative of the frequency-averaged collision integral $\int_\omega F(\omega,\vect{k},t)\delta I(t)/\delta F(\omega,\vect{k},t)$ with respect to $|\vect{k}|$.

\section{Collisionless approximation}\label{sec:MF}
After establishing the memoryless approximation and the conditions under which it satisfies the requirements formulated in Sec.~\ref{sec:general_theory}, we now apply different schemes of dealing with interactions. Specifically, we turn our attention to the explicit evolution of the open $O(N)$-symmetric model at late times as a real-world application. 

At leading order in $1/N$ interactions are described in Hartree approximation. The equations of motion then become (see App.~\ref{sec:2PI})
\begin{align}\label{eq:MF_evolution}
\begin{split}
    \partial_t G^K_\perp&=-\frac{2}{\gamma}\left(D k^2+m_\perp\right)G^K_\perp+\frac{4 i T(D k^2+m_\perp)}{\gamma^2\omega^2+\left(D k^2+m_\perp\right)^2}\\
    \partial_t G^K_\parallel&=-\frac{2}{\gamma}\left(D k^2+m_\parallel\right)G^K_\parallel+\frac{4i T(D k^2+m_\parallel)}{\gamma^2\omega^2+\left(D k^2+m_\parallel\right)^2}\\
    \partial_t \phi&=-\frac{m_\phi}{\gamma} \phi\,,
\end{split}
\end{align}
with $m_\perp=m+\frac{\lambda}{6N}(\phi^2+i(N-1)\int_q G^K_\perp(q)+i\int_q G^K_\parallel(q))$, $m_\parallel=m_\perp+\frac{\lambda}{3N}\phi^2_\text{cl}$ and $m_\phi\equiv m_\perp$ \footnote{Typically the mass $m_\perp$ is negative, which leads to singularities in Eq.~\eqref{eq:MF_evolution}. The problem disappears after integration over frequencies. Later, when including collisions $m_\perp(t)$ will disappear fast enough that we can safely set it to zero in $\Re{G^R}$.}. The only interaction effect enters via the dynamical masses $m_\perp$ and $m_\parallel$, hence, this level of the approximation is collisionless, meaning no process exists that allows excitations to change momentum. In the absence of a non-trivial collision integral, its memoryless approximation is of no concern as all gradient corrections to the mass amount to total derivatives. These vanish for the excess momentum distribution $\delta n(t,\vect{k})=i\int_\omega(G^K(t,\omega,\vect{k})-G^K_\text{th}(t,\omega,\vect{k}))/2$. Its evolution is therefore exact at this order in $1/N$. Integrating over the frequency in Eqs.~\eqref{eq:MF_evolution} one finds a closed set of rate equations for the momentum distributions $n_{\perp/\parallel}(t,\vect{k})$ and the order parameter $\phi(t)$ \footnote{A faster derivation is obtained by directly considering $G^K(t,t,\vect{k})$, which requires no Wigner expansion.}.

The overdamped regime is equivalent to time-dependent stochastic Ginzburg-Landau theory \cite{Abrahams1966,Schmid1966,Gorkov1968,Kapustin2023} or model A in \cite{Hohenberg1977}. The expansion to quadratic order in fluctuations that has previously been used in the context of quench dynamics \cite{Mazenko1985, Dolgirev2020a, Sun2020, Grandi2021} corresponds to the collisionless approximation. It is justified at late times $t\gg 1/\gamma$ if redistribution due to collisions, which will be discussed in the next section, is negligible. As opposed to the closed system analog, the collisionless limit of the open system is not frozen in its initial state as equilibration with the environment persists.

When analyzing the equations Eq.~\eqref{eq:MF_evolution}, we realize that both the order parameter $\phi$ and the longitudinal fluctuations $G^K_\parallel$ are gapped excitations. As a result, they quickly relax the dynamical Goldstone mass $m_\perp(t)$, which therefore plays no role in the evolution at late times. The latter is instead fully determined by the equilibrium decay rate of the excess Goldstone fluctuations. In $d>2$ dimensions one finds
\begin{align}\label{eq:thermal_scaling_MF}
\begin{split}
    \delta G^K_\perp(t,\vect{k})&\approx\left(\delta G^K_\perp(t=0,\vect{k})+2 G^K_{\perp,\text{th}}(\vect{k})\frac{\delta\phi(t)}{\phi_\text{th}}\right)e^{-\frac{2}{\gamma}D\vect{k}^2t}\\
    \delta \phi(t)&\approx -\frac{1}{2\phi_\text{th}}\int_{\vect{k}} \delta G^K_\perp(t,\vect{k})\sim t^{-d/2}\\
    m_\perp(t)&\sim t^{-d/2-1}\,,
\end{split}
\end{align}
where $\delta G^K_\perp(t=0,\vect{k})\sim \vect{k}^0$ represents the excess transversal excitations immediately after the quench. Since in the collisionless approximation the Goldstone theorem is exactly satisfied at all times (see Sec.~\ref{sec:2PI}), the dominant contribution at small momenta $\sim G^K_\text{th}(\vect{k})\sim \vect{k}^{-2}$ is proportional to the displacement of the order parameter $\delta\phi=\phi-\phi_\text{th}$. It therefore decays as $t^{-d/2}$ and can be neglected at late times. 

Based on this, we draw the important conclusion that quenches which preserve the $O(N)$-symmetry differ qualitatively in their late-time behavior from those that do not: In the latter case, Goldstone's theorem does not apply. The excess thermal fluctuations, corresponding to an effectively higher temperature, are then not proportional to $\delta \phi$ and therefore survive for long times, where they are dominant. This also results in a slower equilibration of the order parameter $\delta \phi(t)\sim t^{1-d/2}$ compared to the quench without explicit symmetry breaking. In case of the latter, excess fluctuations scale subthermally. This has an immediate experimental consequence: experiments with perturbations respecting the symmetry, like a temperature quench or Raman scattering, will observe much faster relaxational dynamics than experiments that break the symmetry. For example, the system is expected to recover more slowly from neutron scattering, which induces excitations in the longitudinal mode but not the transverse direction, thereby explicitly breaking the $O(N)$ symmetry \cite{Podolsky2011}. In the next section, we show that collisions will not affect this statement.

\section{Redistribution and screening}\label{sec:Collisions}
In the previous section, we have shown that the collisionless dynamics of open systems at late times reduce to rate equations with dissipative dynamics fully determined by the coupling to the environment. Contrary to the absence of any evolution in collisionless closed systems, it provides insights into the order-parameter dynamics and the density of Gaussian fluctuations. However, we still need to analyze the effect of collisions, the sole origin of thermalization in closed systems. In particular, we will find that collisions render the spectrum in the longitudinal direction gapless. The mutual feedback between longitudinal and transversal dynamics then has to be analyzed carefully to show that the longitudinal evolution adiabatically follows that of the transversal excitations. Although the formalism applies to any dimension $d$ above the lower critical one, the explicit calculations in this section are performed in $d=3$.

Addressing these questions requires an expansion to NLO($1/N$), which describes the same collision processes as in closed systems \cite{Aarts2002,Berges_review_2016}, but with the equations of motion of fluctuations given by Eq.~\eqref{eq:t-evo} instead of Eq.~\eqref{eq:t-evo_closed}. The explicit expressions for the self-energies have been stated in Eqs.~\eqref{eq:2PI_selfenergy} to \eqref{eq:2PI_phi} with the corresponding diagrammatic representation shown in Fig.~\ref{fig:Dyson}.

Near the thermal fixed point, it is useful to subtract the stationary thermal state from the evolution equation, which then becomes
\begin{align}\label{eq:linearKB}
\begin{split}
    Z^{-1}\partial_t \delta G^K=&2i\Re{{G^R}^{-1}}\delta G^K-2i\Re{G^R}\delta\Sigma^K\\&-2i \Re{\delta\Sigma^R}G^K_\text{th}-2i\Re{\delta G^R}\left(\Sigma^K_\text{th}-8i\gamma T\right)\,.
\end{split}
\end{align}
For small perturbations around the thermal state, the last two terms cancel. This excludes however the effect of a finite mass, which for vanishing $\omega$ and $\vect{k}$ is never a small perturbation to $G^R_\text{th}$ and therefore needs to be retained. The combined effect of the second line at small momenta then reproduces the excess population $\sim G^K_\text{th}$ already discussed in the collisionless approximation. Although the dynamical mass is quantitatively modified by collisions as well, the important feature of $O(N)$-symmetric quenches $m_\phi(t)=m_\perp(t)$ ensures that the leading effect of the second line in Eq.~\eqref{eq:linearKB} at small momenta vanishes at late times. 

The qualitatively new effect of collisions is captured by the first line of Eq.~\eqref{eq:linearKB}, with the first term describing the decay of excess fluctuations and scattering of an excitation out of the state with momentum $\vect{k}$. The second term represents the reverse process of collisions into the state on the left-hand side.

An important qualitative difference to the collisionless approximation is that, since the longitudinal mode at long wavelengths can decay into two gapless transverse excitations, it itself becomes gapless in $d<4$. The same is true for the closed $O(N)$ model at finite temperatures \cite{Patashinskii1972, Zwerger2004,Podolsky2011}. In the thermal state in three dimensions one finds \cite{note:bare}
\begin{align}\label{eq:Higgs}
\begin{split}
    G^R_{\parallel,\text{th}}(\omega,\vect{k})&=\frac{1}{2\left(i\gamma\omega-D \vect{k}^2\right)-\phi^2_\text{th}\Pi^R_\text{th}(\omega,\vect{k})}\\
    \Pi^R_\text{th}(\omega,\vect{k})&=\left(\frac{1}{g}+\frac{N T \sin ^{-1}\left(\frac{|\vect{k}|}{\sqrt{2 \vect{k}^2-2 i \frac{\gamma 
   \omega}{D} }}\right)}{8 \pi D^2 |\vect{k}|}\right)^{-1}
\end{split}
\end{align}
and therefore $G^R_{\parallel,\text{th}}\sim \left(\omega+8iD\pi^{-2}\gamma^{-1}\vect{k}^2\right)^{-1/2}$. In other terms, at length scales above the screening length $\xi_\perp=\frac{48 D}{T\lambda}$\cite{note:bare} the formation of transversal fluctuations screens the bare contact interactions.
Although gapless, the spectral weight of the longitudinal modes vanishes compared to that of the transversal excitations for small energies and momenta. Consequently, the density of quench-induced fluctuations vanishes much faster for longitudinal excitations than in the transverse direction. Concluding that the transverse modes are more strongly suppressed at long wavelengths than their longitudinal counterparts, we use the lowest order of the Wigner expansion to write
\begin{align}
    G^K_\parallel(t,\omega,\vect{k})=|G^R_\parallel(t,\omega,\vect{k}) |^2\Sigma^K_\parallel(t,\omega,\vect{k})\,.
\end{align}
Physically, this can be interpreted as the longitudinal modes adiabatically following the slow dynamics of the transverse excitations and preserves the constraints outlined in Sec.~\ref{sec:general_theory}. 

Together with the insight that the order parameter dynamics is relaxing exponentially fast unless driven by the slow dynamics of the transversal mode, we conclude that the longitudinal direction of the order parameter field is always locally equilibrated with the slowly evolving transversal modes. This implies that the interactions between transversal modes according to Eq.~\eqref{eq:Xi}
\begin{align}
    \Xi^R(t,\omega,\vect{k})\approx 2\phi_\text{th}^{-2}\left(i\gamma\omega-D \vect{k}^2\right)
\end{align}
are very strongly screened at distances longer than the longitudinal screening length $\xi_\parallel=\max{\left(\frac{D N T}{16 \phi_\text{th}},\sqrt{\frac{3 D N}{\lambda \phi_\text{th}^2}}\right)}$. On shorter scales, on the other hand, the bare interactions $\Xi^R(t,\omega,\vect{k})\approx 2\lambda/3N$ are recovered. As a result, the long-lived long-wavelength transversal excitations in $\delta G^K$ interact very weakly. At late times one therefore finds that $\delta \Sigma^K\sim (t^{-3/2},\omega^0,\vect{k}^0)$ decays with the same power as the collisionless equilibration with the thermal bath, providing only a small quantitative correction.

Note, that due to $\phi_\text{th}\sim \lambda^{-1/2}$ it is possible to suppress the effects of screening to very long scales by considering weak interaction. However, the smaller overall scattering rate $\sim \lambda^2$ compensates for the less efficient screening, such that the collisions at small momenta cannot be enhanced in this way.

\begin{figure}[tp]
\centering
\includegraphics[width=.95\columnwidth]{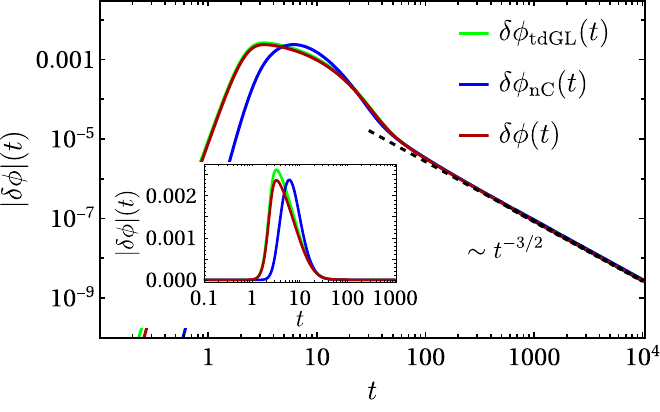}
\includegraphics[width=.95\columnwidth]{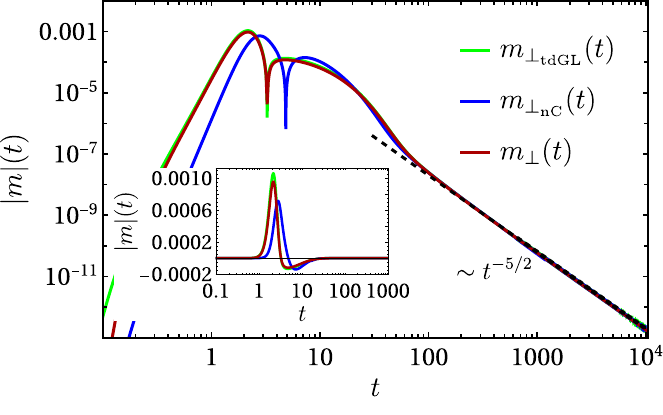}
\includegraphics[width=.95\columnwidth]{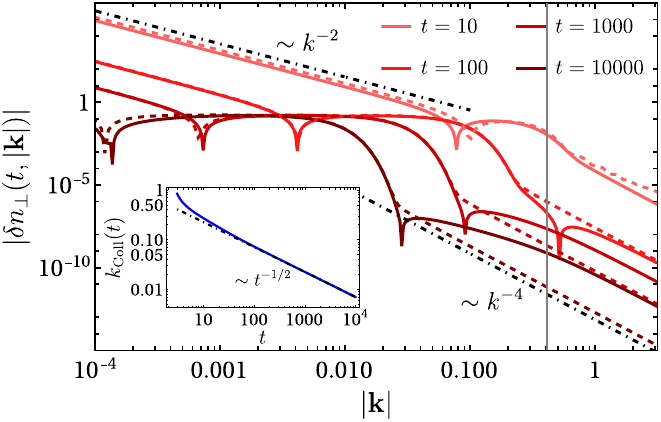}
\caption{Comparison between the time evolution following a weak temperature quench in the collisionless approximation (leading order in $1/N$) and the dynamics including scattering in the memoryless approximation of Eq.~\eqref{eq:KB_Wigner}. Due to the strong screening of interactions at long wavelengths, the asymptotic behavior of the collisionless description in the overdamped approximation (light green) agrees extremely well with the expansion to next-to-leading order in $1/N$ (dark red). Both differ at early times from the collisionless evolution without overdamped approximation (blue), see insets. Both the order parameter (top) and the mass of the transversal mode (middle) follow the prediction Eq.~\eqref{eq:thermal_scaling_MF}. The bottom panel shows the evolution of the transversal momentum distribution. While collisions increase the density at high momenta (full lines) compared to the collisionless (dashed) result, the decay of the thermal peak as well as the momentum scale with the largest number of excitations $k_\text{Coll}$ (see inset) agrees well between the two methods. The inverse screening length $1/\xi_\parallel$ is shown as gray vertical line and we used the parameters $\lambda=T=\gamma=D=1$, $N=10$ and $\phi_\text{th}=\sqrt{5}$.}
\label{fig:quench} 
\end{figure}

We confirm these arguments by an explicit comparison of the dynamics following a small quench of the temperature
\begin{align}\label{eq:quench}
    T(t)=T\left(1+\delta T e^{-\frac{(\gamma t-2)^2}{2\sigma^2}-\frac{1}{\gamma t}}\theta \left(t\right)\right)\,.
\end{align} 
with the quench amplitude set to $\delta T=1/10$ and the pulse width $\sigma=\sqrt{1/5}$. We emphasize that the results at late times are largely independent of the details of the quench protocol. They are reported in Fig.~\ref{fig:quench} with additional results for the longitudinal mode given in App.~\ref{app:longitudinal}.
In Fig.~\ref{fig:quench} the first two panels show the evolution of the order parameter and the transversal mass with the exact, but collisionless evolution shown in blue and the overdamped evolution including collisions and thus redistribution in red. Furthermore, the green line represents the result of Eq.~\eqref{eq:MF_evolution} equivalent to the time-dependent stochastic Ginzburg-Landau equation with weak fluctuations. As expected, due to the strong screening of long-wavelength fluctuations the main difference in these global quantities arises at early times $t\lesssim 1/\gamma$ from the overdamped approximation as is seen in the insets. At late times all methods predict the same behavior independent of collisions and are fully determined by the coupling to the thermal environment. 

The only significant difference that arises from the inclusion of collisions is found at high momenta, where interactions are not screened, see the last panel in Fig.~\ref{fig:quench}. The continued scattering of non-thermal fluctuations from small to large momenta leads to a slower decay of the high momentum tail $\delta n_\perp(t,\vect{k})\sim t^{-3/2}\vect{k}^{-4}$ compared to $\delta n_{\perp,\text{nC}}(t,\vect{k})\sim t^{-5/2}\vect{k}^{-4}$ set by the decay of the transversal mass in the collisionless case. Although the latter results exclusively from the dynamical mass perturbing the thermal background, whereas the former is a consequence of scattering with large momentum transfer, the asymptotic behavior $\sim \vect{k}^{-4}$ applies to both calculations. One also finds that below the inverse screening length $1/\xi_\parallel$ indicated as a gray line in Fig.~\ref{fig:quench}, screening modifies the momentum dependence to $n_\perp(t,\vect{k})\sim \vect{k}^{-2}$. Note, however, that since these excitations at high momenta are short-lived, no significant population can build up. Consequently, the collision integral in both cases is dominated by the same momentum scale $k_\text{Coll}\sim t^{-1/2}$, where the deviation from the thermal state is largest, see inset in Fig.~\ref{fig:quench}. 

We can therefore conclude that, due to the efficient screening of interactions in the ordered state of the $O(N)$-model, the order parameter excitations never thermalize to a temperature different from that of the environment. There is thus no extended transient regime, where the order parameter fluctuations have thermalized but not yet equilibrated with the environment. In fact, the excess fluctuations at late times possess a subthermal momentum distribution, wherefore the three-temperature model should not be applied there. In particular, the 3-TM predicts a qualitatively wrong asymptotic decay of the order parameter displacement. We expect these results to generalize as follows: for a weak coupling fixed point, no thermalization occurs, as relaxation eradicates the quench-induced population before it thermalizes. For a strong coupling fixed point, on the other hand, one expects the opposite behavior and hence a three-temperature model to be justified after the thermalization time.

\section{open-system Boltzmann equation}\label{sec:Boltzmann}
For simulations and the evaluation of observables, it is convenient to work directly with the equations of motion of the retarded and Keldysh Green's function. The structure of the collision terms, however, is most clearly revealed in the kinetic equation obtained by a rephrasing of the equations of motion in terms of the distribution function $F=i G^K/\mathcal{A}$, where $\mathcal{A}=-2\Im{G^R}$ denotes the spectral function. In equilibrium $F$ is independent of momentum and related to the Bose distribution $n_B(\omega)$ by $F_\text{th}(\omega)=2n_B(\omega)+1=\coth{(\omega/(2T))}$. We begin with the observation that $\partial_t G^K\approx -i\mathcal{A}\partial_t F$ since the relevant perturbations to the spectral function at small momenta, for example $m_\perp\sim t^{-5/2}$, vanish quickly. Technically this is justified because the collisions in $\Sigma^R$ at most contain one less distribution function than the corresponding term in $\Sigma^K$. As a result, higher momenta contribute to $\Sigma^R$, which therefore relaxes more quickly.

The kinetic equation then takes the form
\begin{align}
\begin{split}
    Z^{-1}i\partial_t F(\omega,\vect{k})=&\frac{4D\vect{k}^2+\Sigma^R+\Sigma^A}{-i\left(\Sigma^R-\Sigma^A-4i\gamma\omega\right)}\\&\times\!\left[8\gamma T\!+i\Sigma^K\!-i\left(\Sigma^R\!-\!\Sigma^A\!-4i\gamma\omega\right)\!F\right],
\end{split}
\end{align}
with the spectral function at small arguments given by
\begin{align}
    \mathcal{A}(\omega,\vect{k})=\frac{1}{\gamma}\frac{\omega}{\omega^2+\epsilon(\vect{k})^2}\,,
\end{align}
where $\epsilon(\vect{k})=(D\vect{k}^2+m_\perp)/\gamma$.
On the other hand, the Keldysh self-energy and thermal term from the bath approach a constant. Hence the frequency dependence of $G^K$ mimics that of $G^R+G^A$. Together with $(G^R+G^A)/\mathcal{A}\sim \epsilon(\vect{k})/\omega$ this implies for small arguments the form
\begin{align}
F(\omega,\vect{k})\approx 2f(\vect{k}) \epsilon(\vect{k})/\omega\equiv 2 T(\vect{k})/\omega    
\end{align}
with the dimensionless momentum distribution function $f(\vect{k})$. Physically this can be interpreted as a distribution with a momentum-dependent temperature $T(\vect{k})$, corresponding to the assumption of local thermal equilibrium in momentum space. If this representation is valid on the scale set by $k_\text{Coll}(t)$, the long-lived modes at small momenta allow one to simplify integrals of the form
\begin{align}
    \int_\omega \mathcal{A}(\omega,\vect{k})F(\omega,\vect{k})C(\omega,\vect{k})\approx \frac{1}{\gamma}f(\vect{k})C(0,\vect{k})
\end{align}
for a smooth function $C(\omega,\vect{k})$ varying slowly on the energy scale $\epsilon(\vect{k})$. 

With this, the kinetic equation can be projected to the zero-energy modes, where it takes the form of a Boltzmann equation
\begin{align}\label{eq:Boltzmann}
    \partial_t f(\vect{k})=2 T-2 \epsilon(\vect{k})f(\vect{k})+\frac{1}{\gamma^4}I_\text{coll}[f](\vect{k})\,.%2\gamma\partial_t f=-iZ^{-1}\partial_t f\,.
\end{align}
within the collision integral
\begin{align}\label{eq:Collision}
\begin{split}
I_\text{coll}[f](\vect{k})&=\!\int_{\vect{q}}\int_{\vect{p}}\Omega(\vect{k},\vect{q},\vect{p})\left\{\epsilon(\vect{k})f(\vect{k})\right. \\&\left.\times\left[f(\vect{p})f(\vect{k}-\vect{q})\!+\!\left(f(\vect{p})\!+\!f(\vect{k}-\vect{q})\right)f(\vect{q}-\vect{p})\right]\right.\\&\left.-f(\vect{p})f(\vect{k}-\vect{q})f(\vect{q}-\vect{p})\right.\\&\left.\times\left[\epsilon(\vect{p})+\epsilon(\vect{k}-\vect{q})+\epsilon(\vect{q}-\vect{p})\right]\right\}\,,
%    I_\text{coll}[f](\vect{k})=&\!\int_{\vect{q}}\int_{\vect{p}}\Omega(\vect{k},\vect{q},\vect{p})\left[f(\vect{k}-\vect{q})f(\vect{p})f(\vect{q}-\vect{p})\right.\\&\left.-f(\vect{k})\left(f(\vect{k}-\vect{q})f(\vect{q}-\vect{p})-f(\vect{k}-\vect{q})f(\vect{p})\right.\right.\\&\qquad\quad\left.\left.+f(\vect{q}-\vect{p})f(\vect{p})\right)\right]\,.
\end{split}
\end{align}
where, assuming $f(\vect{k})\gg 1$, we have retained only terms with the highest power in $f$.
As per the discussion below Eq.~\eqref{eq:Higgs}, we have furthermore exploited that the longitudinal degrees of freedom adiabatically follow the slow transversal modes and therefore do not contribute to the scattering. In Eq.~\ref{eq:Boltzmann} the first term inserts thermal fluctuations, which decay due to the second term. The rate of collisions is determined by the on-shell scattering matrix $\Xi^R(\omega=0,\vect{q})$ via
\begin{align}
\begin{split}
    \Omega(\vect{k},\vect{q},\vect{p})=&\frac{1}{\epsilon(\vect{k})+\epsilon(\vect{p})+\epsilon(\vect{k}-\vect{q})+\epsilon(\vect{q}-\vect{p})}\\&\times\frac{1}{\epsilon(\vect{p})+\epsilon(\vect{k}-\vect{q})+\epsilon(\vect{q}-\vect{p})}|\Xi^R(0,\vect{q})|^2\,,
\end{split}
\end{align}
which in our case is heavily screened at small momenta since $|\Xi^R(\omega=0,\vect{q})|^2\approx 4 \phi_\text{th}^{-4} D^2 \vect{q}^4$.
In Eq.~\eqref{eq:Collision} the first line describes events scattering out of the momentum mode $\vect{k}$, whereas the second one scatters excitations into that mode. It is easy to check that the thermal distribution in low-frequency approximation $f(\vect{k})=T/\epsilon(\vect{k})$ is a stationary solution to the Boltzmann equation, as it nullifies the collision integral.

The open-system Boltzmann equation derived here is numerically no simpler to evaluate than the gradient expansion of the Kadanoff Baym equation. It is however valuable for scaling arguments analogously to its closed system relative \cite{Micha2004, Berges2008, Berges2011}.
For the discussion of the thermal fixed point in the spontaneously symmetry-broken phase, we have already seen that the collision integral provides only a subleading correction. The same conclusion is recovered from simple scaling arguments in Eq.~\eqref{eq:Boltzmann}. Consequently, a more detailed analysis is dispensable in the current context and should be deferred to a system with stronger interactions at long wavelengths.

We emphasize that the open-system Boltzmann equation is valid for late
times, where all occupied modes are overdamped. Energy conservation
therefore enforces a very different constraint from the usual case,
where the dispersion relation constraints the collision integral
\cite{Wais2018, Picano2021, Kamenev_book}.

\section{Outlook}\label{sec:Outlook}
In this paper, we have established a framework for the theory of open system dynamics at late times as a hierarchy of successive simplifications. In doing so we have included an investigation of the conditions necessary for their applicability. Implementing it on the open $O(N)$-symmetric model of an $N$-component order parameter and its fluctuations in the symmetry-broken phase, we have shown that near the thermal fixed point momentum redistributing collisions are irrelevant, resulting in a collisionless evolution. The latter is fully described in terms of the dynamical mass of the transverse modes determining the effective rate of their equilibration with the environment as well as the order parameter relaxation. The resulting Gaussian scaling of the momentum distribution is dominated by a time-dependent momentum scale $k_\text{Coll}\sim t^{-1/2}$, which pushes the Wigner expansion, used in the derivation, to the limit of its range of validity. Nevertheless, the expansion remains justified by the strong screening of interactions at long wavelengths for near-thermal states. Hence, although observables that require a frequency resolution exceeding $1/t$ are not captured, collisions are described correctly. Consequently, predictions for the late-time evolution of the $O(N)$ model are reliable.

One can interpret the collisionless evolution as the dynamical manifestation of a Gaussian fixed point in an overdamped system. The resulting collisionless dynamics, controlled by the mass as the sole remaining dynamical parameter, is hence expected to be universal, applying to other overdamped systems with de facto weak infrared interactions. As opposed to the evolution of closed systems, where thermalization requires momentum redistribution, this admits the efficient simulation of the time evolution of a large class of open systems. 

To explicitly check which systems lie in this class can be rather demanding. However, a good indication regarding the applicability of the collisionless evolution can be conveniently obtained from the open-system Boltzmann equation \eqref{eq:Boltzmann}. Using the distribution function $f(\vect{k})$ from the collisionless evolution, one has to compare the long-wavelength limit of the collision integral and the dissipation rate, i.e. $\lim_{|\vect{k}|\to 0}I_\text{coll}[f](\vect{k})/(\epsilon(\vect{k})f(\vect{k}))$. If it vanishes, momentum redistribution has no impact on the late-time evolution, and the collisionless description is considered justified.

This leaves the question about other systems with less efficient screening, for example, in lower dimensions. Even more interestingly, there is no need to constrain oneself to the thermal fixed point. The road map developed here can be applied in the vicinity of non-thermal fixed points accessible in strong quenches for which indications have already be seen in experiments \cite{Vogelgesang2018,Zhou2021, delaTorre2022}. The highly non-thermal distributions realized for a strong quench are generally expected to lead to weaker screening and therefore require a separate discussion. It will be interesting to see, to which extent the hierarchy of simplifications developed here for near-thermal states can be applied there and how non-trivial critical exponents can be recovered or predicted. It should however be noted, that for condensed matter experiments $N\leq 3$. As a result, strong quenches to the disordered phase can generate topological defects, which may require a separate discussion.
%\jlc{A strong quench results in a gapped spectrum at short times that wipes out the thermal background. Mean-field at short times then shows that all momenta get occupied at the same rate due to the constant coupling to the bath. The resulting subthermal momentum distribution screens less efficiently as RPA-bubbles at small energies are much less Bose enhanced.}

Note added: After the publication of our work, an independent study of the time evolution of spin glasses appeared \cite{Hosseinabadi2023a} that also applies the 2-PI effective action formalism to open systems \cite{Hosseinabadi2023b}.

\section{acknowledgements}
The work was supported by the Deutsche Forschungsgemeinschaft (DFG, German Research Foundation) CRC 1238 project number 277146847. MB acknowledges support from the Deutsche Forschungsgemeinschaft (DFG, German Research Foundation) under Germany's Excellence Strategy Cluster of Excellence Matter and Light for Quantum Computing (ML4Q) EXC 2004/1 390534769.

\appendix
%\section{Effective action}\label{app:EffAct}
%In the interests of a self-contained presentation, this appendix provides a brief summary of one-particle irreducible (1PI) and two-particle irreducible (2PI) effective actions. Comprehensive introductions to the topic are found in the literature both in equilibrium \cite{Amit1984} and out of equilibrium \cite{Berges_review_2016}.

%As we argue in the main text it is essential that our theory is conserving. However, targeting the symmetry-broken phase, we also require our approximation to produce gapless Goldstone modes, as otherwise all late-time behavior will be cut off by the time scale imposed by the erroneous Goldstone mass. We will thus proceed to estimate the mass of the Goldstone mode obtained from 1PI and 2PI effective actions.

\section{Derivation of the 1PI effective action}\label{app:1PI}

In this appendix, we provide the detailed derivation of the one-particle-irreducible effective action and the Green's functions derived from it. The emphasis is put on the recovery of Ward-Takahashi identities and thereby Goldstone's theorem on the one hand and the violation of conservation laws on the other hand.

\subsection{Leading order in $1/N$}

The 1PI effective action \eqref{eq:1PI-action} has been stated in the main text. We begin its evaluation by noting that the classical action $S[\phi]$ is given in Eq.~\eqref{eq:action} and the one-loop term reads
\begin{align}
\begin{split}
    \frac{i}{2}\Tr_\mathcal{C}\ln{G_0^{-1}(\phi)}=&\frac{i}{2}\ln \det\begin{pmatrix}
    \left[G_{0}^{-1}\right]^V&\left[G_{0}^{-1}\right]^A\\
    \left[G_{0}^{-1}\right]^R&\left[G_{0}^{-1}\right]^K
    \end{pmatrix}\,.
\end{split}
\end{align}
Similar to the trace, the determinant is also taken over space-time, Keldysh index and field components (running from 1 to $N$), the latter of which we will denote with Greek letters. Since in the following we will need to take derivatives with respect to the fields, we have to write the inverse bare Green's function components without the restriction to the physical field configuration $\phi_{\alpha,\text{cl}}\sim \delta_{1\alpha}$ and $\phi_{\alpha,\text{q}}=0$. Using Einstein notation, we find
\onecolumngrid
\begin{align}\label{eq:1-PI_bareG}\small
\begin{split}
    \left[G_{0}^{-1}(x,y,\phi)\right]^{R/A}_{\alpha\beta}=&\delta^{(4)}(x-y)\left\{\left[2(-\partial_t\mp i\gamma\partial_t-m+D \vect{\nabla}^2)-\frac{\lambda}{3N}\sum_{\sigma\in\{\text{cl},\text{q}\}}\phi^2_{\gamma,\sigma}(x)\right]\delta_{\alpha\beta}-\frac{2\lambda}{3N}\sum_{\sigma\in\{\text{cl},\text{q}\}}\phi_{\alpha,\sigma}(x)\phi_{\beta,\sigma}(x)\right\}\\
    \left[G_{0}^{-1}(x,y,\phi)\right]^{K}_{\alpha\beta}=&\delta^{(4)}(x-y)\left\{\left(8i\gamma T-\frac{2\lambda}{3N}\phi_{\gamma,\text{cl}}(x)\phi_{\gamma,\text{q}}(x)\right)\delta_{\alpha\beta}-\frac{2\lambda}{3N}\sum_{\sigma\in\{\text{cl},\text{q}\}}\phi_{\alpha,\sigma}(x)\phi_{\beta,\bar{\sigma}}(x)\right\}\\
    \left[G_{0}^{-1}(x,y,\phi)\right]^{V}_{\alpha\beta}=&-\delta^{(4)}(x-y)\left\{\frac{2\lambda}{3N}\left(\phi_{\gamma,\text{cl}}(x)\phi_{\gamma,\text{q}}\delta_{\alpha\beta}(x)+\sum_{\sigma\in\{\text{cl},\text{q}\}}\phi_{\alpha,\sigma}(x)\phi_{\beta,\bar{\sigma}}(x)\right)\right\}\,,
\end{split}
\end{align}
\twocolumngrid
which contains terms of order $N^0$ that are diagonal in field index, as well as contributions of order $N^{-1}$. While discussing the effective action at leading order in $1/N$ we may safely neglect the latter. They will however be included in our discussion beyond the leading order. Restricting these equations to the classical physical field configuration leads to Eq.~\eqref{eq:Gphi}.

\begin{figure}[ht]
\centering
\includegraphics[width=.9\columnwidth]{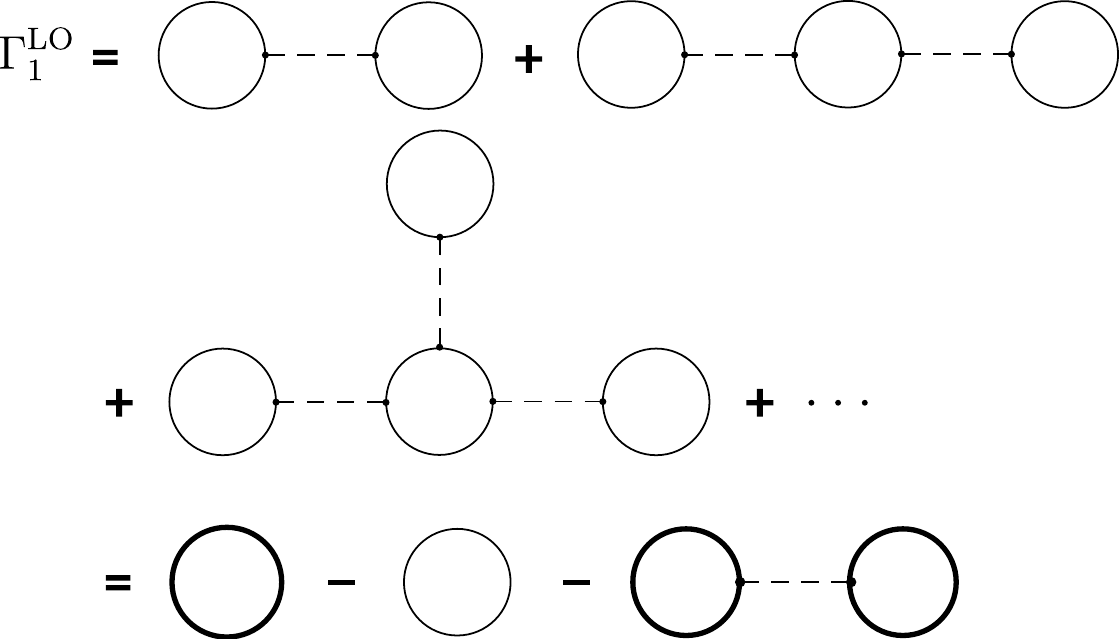}
\caption{The leading order of $\Gamma_1$ in the expansion in $1/N$ consists of all tadpole diagrams with at least two loops. Note, that the bare Green's function (thin line) already contains the coupling to the field $\phi$ (see Eq.~\eqref{eq:1-PI_bareG}). In addition, the dressed Green's functions (thick lines) contain self-energy corrections in the form of tadpole diagrams. The thin dashed line indicates the bare vertex. Symmetry factors and Keldysh structure have been suppressed.} 
\label{fig:Gamma1LO}
\end{figure}

The interaction part $\Gamma_1[\phi]$ contains all one-particle irreducible diagrams that involve at least two loops. To leading order in $1/N$ the one-particle irreducible action includes all tadpole diagrams, see Fig.~\ref{fig:Gamma1LO}. Carefully accounting for the symmetry factors one realizes that these may be collected in the closed expression \cite{Amit1984}
\begin{align}
\begin{split}
    \Gamma_1^\text{LO}[\phi]=&\frac{i}{2}\Tr_\mathcal{C}\ln{G_{\text{LO}}^{-1}(\phi)}-\frac{i}{2}\Tr_\mathcal{C}\ln{G_{0}^{-1}(\phi)}\\&-\frac{\lambda}{6N}\Tr_\mathcal{C}\left(G_\text{LO}(x,x,\phi)\cdot\tau^1\cdot G_\text{LO}(x,x,\phi)\right)\,.
\end{split}
\end{align}
where $G_{\text{LO}}$ denotes the Green's function in Hartree approximation
\begin{align}
\begin{split}
    \left[G_{\text{LO}}^{-1}(\phi)\right]_{\alpha\beta}=&\left[G_{0}^{-1}(\phi)\right]_{\alpha\beta}-\Sigma_{\text{H},\alpha\beta}(\phi)\\
    \Sigma_{\text{H},\alpha\beta}(x,x,\phi)=&\frac{i\lambda}{3N} \Tr_f\left(\left\{G_{\text{LO}}(x,x,\phi), \tau^1\right\}_+\right)\delta_{\alpha\beta}\,,
\end{split}
\end{align}
with $\{\cdot,\cdot\}_+$ denoting the anti-commutator and the trace being performed over field components and space-time but not the Keldysh index. Furthermore, contributions $\sim N^{-1}$ are neglected in the bare Green's function. The projection to the physical field configuration results in Eq.~\eqref{eq:GHartree}.
The Keldysh structure of the bare vertex is contained in $\tau^1=\mathbb{1}_{N\times N}\otimes \sigma^1$, which acts as the identity matrix on the space of field components and as Pauli matrix $\sigma^1$ in Keldysh space.

To find the equations of motion for the field expectation value and the transversal mode, we need to take derivatives with respect to $\phi_{\text{q},\text{cl}}$. We begin with
\begin{align}\label{eq:EOM_LO}
    \frac{\delta\Gamma^{\text{LO}}[\phi]}{\delta \phi_{\alpha,\text{q}}(x)}\bigg|_{\phi_\text{cl}=\phi=\langle\hat{\phi}\rangle}\overset{!}{=}0\,,
\end{align}
which is in general hard to evaluate because derivatives can act on the Hartree self-energy $\Sigma_\text{H}$ resulting in an infinite series of chain rules. Fortunately, all derivatives acting on $\Sigma_\text{H}$ cancel in Eq.~\eqref{eq:EOM_LO} leaving us with the simple expression
\begin{align}
    \frac{\delta\Gamma^{\text{LO}}_1[\phi]}{\delta \phi_{\alpha,\text{q}}(x)}\bigg|_{\phi_\text{q}=0}=-\frac{i\lambda}{3N}\phi_{\beta,\text{cl}}(x)\Tr_\mathcal{C} \left(G_\text{LO}(\phi)-G_0(\phi)\right)(x)\,.
\end{align}
Together with the zero- and one-loop contributions and using $\phi_{\alpha,\text{cl}}= \phi\delta_{1\alpha}$ this combines to the equation of motion for the order parameter~\eqref{eq:1PI_op}.

Following the same procedure for the second functional derivative, one finds the self-energies stated in Eq.~\eqref{eq:Sigma1PI}.

\subsection{Next-to-leading order in $1/N$}\label{app:1PI-NLO}

\begin{figure}
\centering
\includegraphics[width=.9\columnwidth]{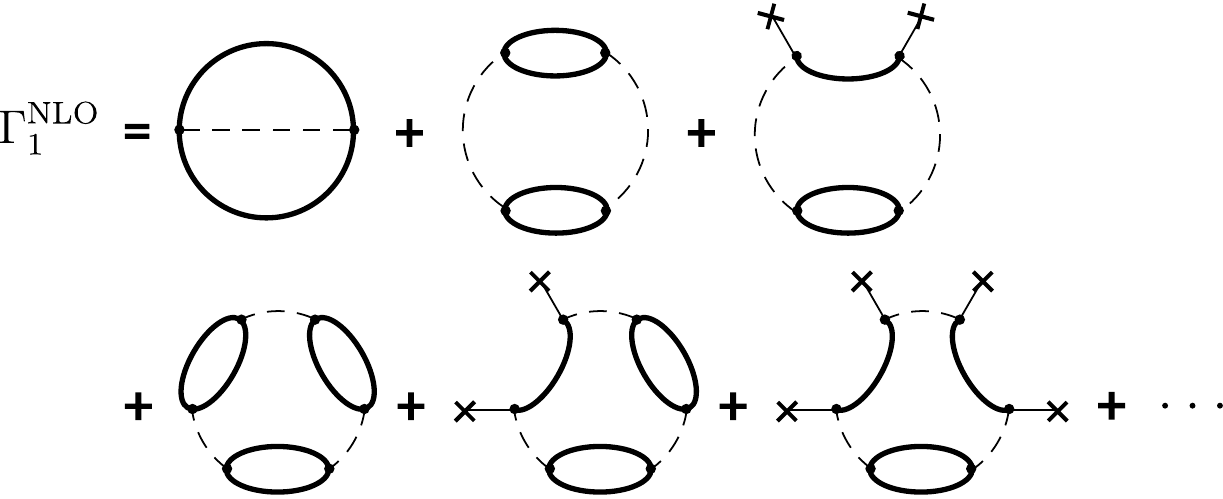}
\caption{The diagrammatic representation of $\Gamma_1$ up to NLO($1/N$) is the sum of all interaction rings with any number of screening bubbles inserted. Note, that each bubble may be opened and replaced by fields as long as the final diagram contains at least two loops. The corresponding one-loop diagrams are already included in $\Tr_\mathcal{C}\ln G^{-1}$. Here we show the first three terms of this series.} 
\label{fig:Gamma1NLO}
\end{figure}

The additional Feynman diagrams contributing to $\Gamma_1[\phi]$ at next-to-leading order are shown in Fig.~\ref{fig:Gamma1NLO}. Note that diagrammatically $\Gamma_1^\text{NLO}$ is very similar to its 2PI counterpart $\Gamma_2^\text{NLO}$, with the crucial difference that in the former any number of Green's functions can be replaced by fields as long as the expression remains one-particle irreducible and consists of at least two loops. The related one-loop diagrams are then included via $\Tr_\mathcal{C}\ln{G_0^{-1}(\phi)}$, where, as opposed to the previous discussion in leading order in $1/N$, the terms $\sim 1/N$ have to be kept. Adding both to the effective action, we find
\begin{align}
   \Gamma^\text{NLO}=\Gamma^\text{LO}+\frac{i}{2}\Tr_\mathcal{C} \ln\left(1-\frac{2\lambda}{3 N}\left[P(G_\text{LO}+\phi\times \phi)\right]\right)
\end{align}
with $\phi \times \phi$ denoting the dyadic product in Keldysh space and the screening bubble abbreviated by
\begin{align}
    P(X)=\frac{i}{2}\begin{pmatrix}
        \Tr(X\cdot X^\top)&\Tr(X\cdot \sigma^1 \cdot X^\top)\\
        \Tr(X \cdot X^\top\cdot \sigma^1) & \Tr(X\cdot \sigma^1 \cdot X^\top\cdot \sigma^1)
    \end{pmatrix}\,.
\end{align}
Here products are local in space-time and field index and the matrix notation refers exclusively to the Keldysh structure.
The equation of motion for the classical field is once again obtained from the functional derivative of the effective action. Upon reinstating the Keldysh structure and assuming a homogeneous stationary state it can be written as
\onecolumngrid
\begin{align}\label{eq:EOM_1-PI_NLO}
\begin{split}
    0\overset{!}{=}\frac{\delta \Gamma^\text{NLO}}{\delta \phi_q}=
    &\left\{\omega^2+i\gamma\omega-m-D \vect{k}^2-\frac{\lambda}{6N}\left(\phi^2+i\int_q G^K(q)\right)\right.-\frac{i}{2}\left(W^R\star G^K_\perp+W^K\star G^R_\perp\right)(k)\\
    &+i\left(\frac{\lambda}{3N}\right)^2\Pi^R(0)\left[\int_{q}\left(|G^R(q)|^2B^K(q)+2G^K(q)\Re{\left(B^R(q)G^R(q)\right)}\right)\right.\\
    &\left.\left.\qquad\qquad\qquad\qquad\;+i\phi^2\int_q\left(|G^R_\parallel(q)|^2 W^K(q)+2G^K_\parallel(q)\Re{\left(G^R_\parallel(q)W^R(q)\right)}\right)\right]\right\}\phi\,,
\end{split}
\end{align}
where
\begin{align}
    W^R(k)=&\frac{1}{\frac{3N}{2\lambda}-i(G^K\star G^R)(k)-\phi^2 G^R_\parallel(k)}\\
    W^K(k)=&|W^R(k)|^2\left[\phi^2 G^K_\parallel(k)+\frac{i}{2} \left(G^K\star G^K+2\Re{\left(G^R\star G^R\right)}\right)(k)\right]\\
    B^R(k)=&\left(W^K\star G^R+W^R\star G^K\right)(k)\\
    B^A(k)=&\left(W^K\star G^A+W^A\star G^K\right)(k)\\
    B^K(k)=&\left(2\Re{\left(W^R\star G^R\right)}+W^K\star G^K\right)(k)\,,
\end{align}
\twocolumngrid
\noindent
and all Green's functions are to be interpreted as $G_\text{LO}$.
Since the transversal Green's function is not allowed to carry any uncontracted $\phi$ fields, a derivative with respect to the classical field can act only on the field outside the bracket in Eq.~\eqref{eq:EOM_1-PI_NLO}. As a consequence one finds Eq.~\eqref{eq:Goldstone1PI}. 

\begin{figure}[h!]
\centering
\includegraphics[width=.9\columnwidth]{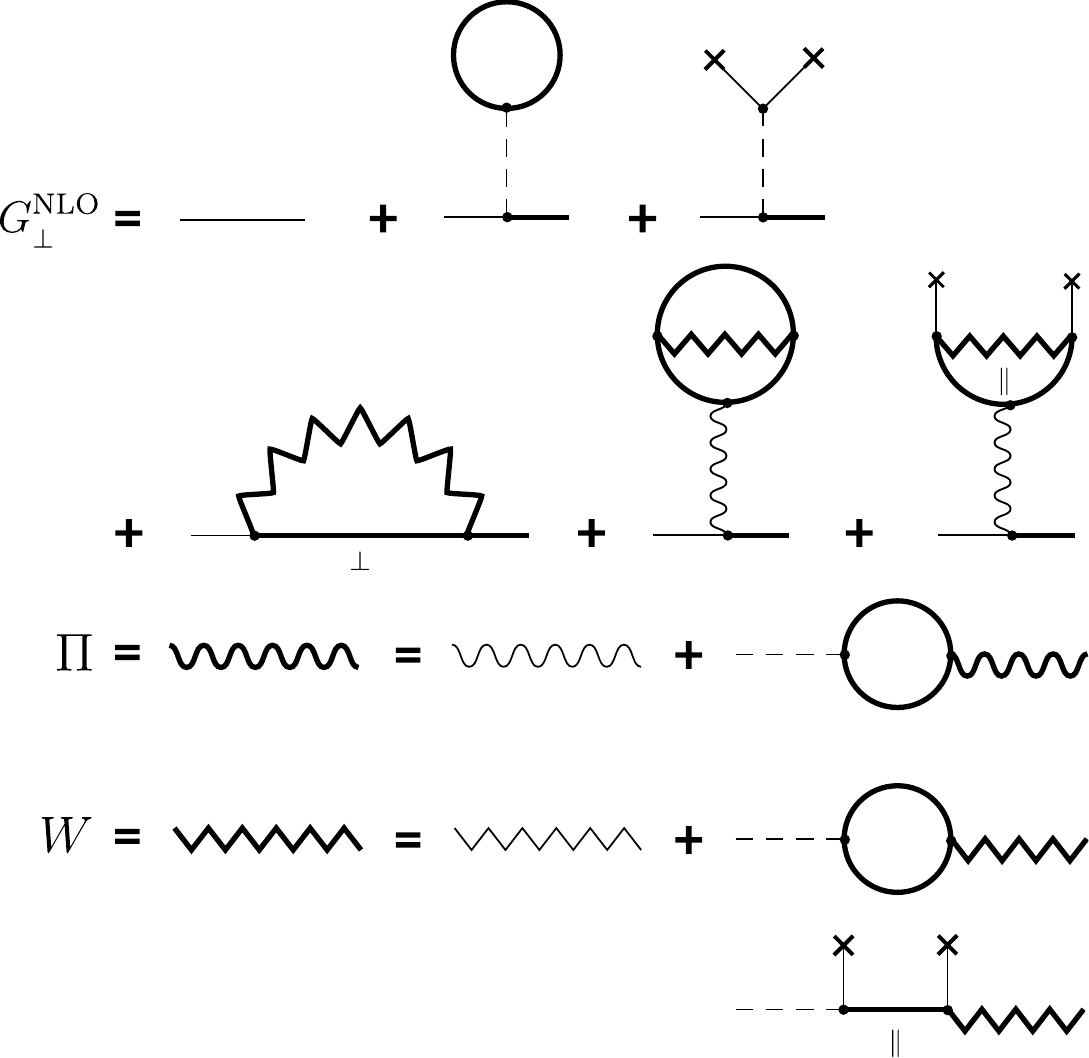}
\caption{Diagrammatic representation of the 1PI transversal Green's function $G_\perp$ in next-to-leading order in $1/N$. For notational simplicity, we have suppressed the Keldysh structure. One may check that the self-energy is the sum of all one-particle irreducible diagrams that are no smaller than $1/N$. In particular, we point out that the restriction to irreducible self-energies implies a difference between the dressed Hartree interaction $\Pi$ and the dressed Fock vertex $W$. By making all interactions explicit to ease the identification of diagrams in Eq.~\eqref{eq:EOM_1-PI_NLO} we break with our previous convention as here the bare Green's function lacks interactions with the field (last diagram in the first line).} 
\label{fig:1-PI_NLO}
\end{figure}

While the expression in Eq.~\eqref{eq:EOM_1-PI_NLO} is bloated by the Keldysh structure, it nevertheless has a relatively simple diagrammatic interpretation shown in Fig.~\ref{fig:1-PI_NLO}. There we identify the first line of Eq.~\eqref{eq:EOM_1-PI_NLO} with the first four diagrams contributing to $G^\text{NLO}_\perp$ and the remaining two lines with the other diagrams whose order is the same in both representations.

We spare ourselves the computation of the longitudinal propagator, which is dressed by many more self-energy insertions, and instead contend ourselves with noting that there will again be processes that describe the decay or creation of longitudinal excitations from pairs of transversal ones without the corresponding term in the self-energy of $G_\perp$. As opposed to the calculation at leading order, the first diagram in the second line in Fig.~\ref{fig:1-PI_NLO} now balances the decay of a longitudinal excitation into two transversal ones. There are however additional processes, suppressed as $1/N^2$ for the transversal mode that are missing from the longitudinal mode at the current expansion order. Hence, at time-scales $t \gtrsim N^2\tau$ violations of conservation laws can reach $\mathcal{O}(1)$. Since these considerations also apply to the large background of thermal fluctuations, it is preferential to evade these issues and instead focus on conserving approximations.

\section{Wigner transform}\label{app:Wigner}

A very intuitive and physically insightful way of dealing with convolutions
between two-time functions $C(t_1,t_2)=\int_{-\infty}^\infty dt_3 A(t_1,t_3)B(t_3,t_2)\equiv A\circ B$ 
that arise in the context of time evolution, 
is obtained in terms of their Wigner transform \cite{Kamenev_book}
\begin{align}
    A(t,\omega)=\int_{-\infty}^\infty d\tau e^{i\omega \tau}A\left(t+\frac{\tau}{2},t-\frac{\tau}{2}\right)
\end{align}
with inverse transformation
\begin{align}
    A(t_1,t_2)=\int_{-\infty}^\infty \frac{d\omega}{2\pi} A(t,\omega)\,,
\end{align}
where $t=\frac{1}{2}\left(t_1+t_2\right)$ and $\tau=t_1-t_2$. The Wigner
transformation of a convolution $C(t_1,t_2)$
is then given by an infinite series of derivatives
\begin{align}\label{eq:WignerSeries}
    C(t,\omega)=A(t,\omega)e^{\frac{i}{2}\left(\cev{\partial}_\omega\vec{\partial}_t-\cev{\partial}_t\vec{\partial}_\omega\right)}B(t,\omega)\,,
\end{align}
where the arrows indicate the direction of the differentiation.
Clearly, in the time-translation invariant case correlation functions depend only on
time differences and one recovers the well-known convolution theorem
\begin{align}
    C(\omega)=A(\omega)B(\omega)\,.
\end{align}
Hence, for slowly evolving systems a systematic expansion around the stationary
limit arises. Furthermore, if the series can be truncated, the time nonlocal
convolution has been replaced by an instantaneous term. Physically, this means
that each memory integral can be replaced by a set of coupled Markovian terms,
which allows one to make use of the large and often highly efficient toolkit developed
for Markovian systems.

These useful properties, make the Wigner expansion very popular. However, the crux
lies in the convergence properties of the series of derivatives in Eq.~\eqref{eq:WignerSeries}.
It is frequently argued, that the Wigner approximation is valid for systems with 
well-defined quasiparticles, since the distribution function varies on scales $\delta_\omega\approx T$
and evolution takes place on scales of the quasiparticle lifetime $\delta\tau\approx \tau_\text{qp}$.
The resulting slow variation inequality $\delta_\omega\delta_\tau\gg 1$ is
then indeed equivalent to the condition for well-defined quasiparticles \cite{Kamenev_book}.

There is however a flaw in this simple argument: While it is true that the
distribution function $F(t,\omega)$ defined by $G^K=G^R\circ F - F\circ G^A$ 
varies slowly in frequency, the same is not true for the Keldysh Green's function. 
It is the latter, however, that enters into the collision integral. Consequently 
the argument breaks down beyond linear-response theory, since then the dynamical 
Keldysh Green's function enters the collision integral self-consistently. We
can therefore trace the breakdown of the Wigner expansion to terms of the form
$\partial_\omega G^R(t,\omega)\partial_t F(t,\omega)\sim 1$. Hence any system
evolving on a time-scale $\sim \tau_\text{qp}$ limits the controlled use of 
the Wigner expansion to linear response. Nevertheless, numerically it has been
demonstrated that the Wigner expansion can still be a good approximation at late 
times \cite{Berges2006}.

Note the explicit reliance of the above argument on the existence of a time-scale $\tau_\text{qp}$.
For critical and self-similar dynamics no such scale exists and the argument
needs to be reconsidered. At first sight, scaling dynamics implies an algebraic 
relaxation $\sim t^\alpha$, which means that all gradient corrections can be neglected 
at sufficiently late times. However, this is only the case if the frequency scale of
the Keldysh Green's function does not depend on time, which in general it does. This topic 
will be discussed in detail in Sec.~\ref{sec:Gapless}.

\section{Dynamics of the longitudinal mode}\label{app:longitudinal}
\begin{figure}[tp]
\centering
\includegraphics[width=.95\columnwidth]{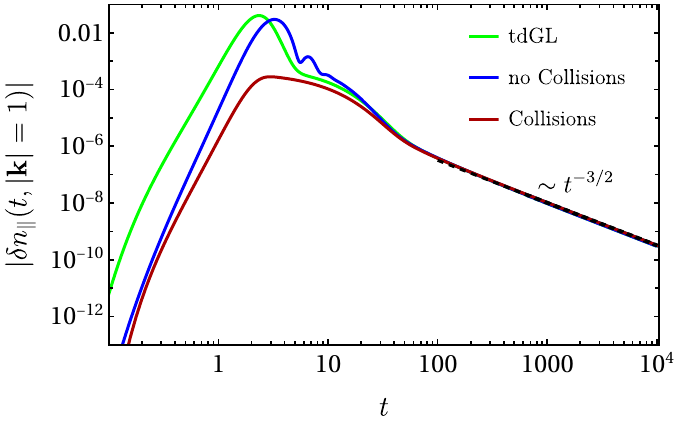}
\includegraphics[width=.95\columnwidth]{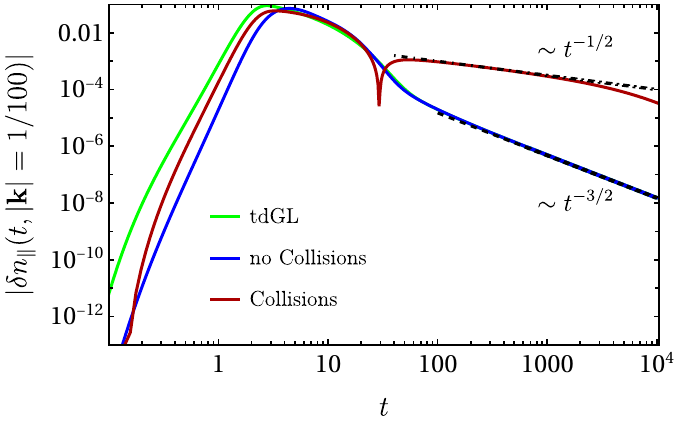}
\caption{Time evolution of the transverse mode following the weak temperature quench discussed in the main text. The comparison shows the collisionless approximation (leading order in $1/N$) (light green), the collisionless evolution without overdamped approximation (blue), and the scattering dynamics in the memoryless approximation of Eq.~\eqref{eq:KB_Wigner} (dark red). At momenta larger than the inverse screening length, $k> 1/xi_\parallel$ (top panel), all methods agree at late times. At smaller momenta $k \ll 1/\xi_\parallel$, however, the effect of collisions dominates at late times (bottom panel). The same parameters as in the main text are used $\lambda=T=\gamma=D=1$, $N=10$ and $\phi_\text{th}=\sqrt{5}$ corresponding to an inverse screening length $1/\xi\approx 0.4$.}
\label{fig:transverse} 
\end{figure}

To complement the results discussed in Sec.~\ref{sec:Collisions}, we now discuss the corresponding dynamics of the longitudinal mode. As is discussed in the main text, the gap of the longitudinal mode is closed in the 2PI formulation by interactions at next-to-leading order in $1/N$. It is at this order of the expansion, that the decay of the longitudinal mode into two transversal modes is included. Consequently, one expects a more significant correction relative to the results of the time-dependent Ginzburg-Landau theory. Due to the small total density of these transverse excitations, however, this does not affect the overall evolution of the system. To illustrate this, we show the density of the longitudinal mode at high and low momenta in Fig.~\ref{fig:transverse}. For the higher momentum $|\mathbf{k}|> 1/\xi_\parallel$ and late times, the transverse mode is well described by the bare propagator with dynamical mass $m_\parallel(t)$. Consequently, we find the same decay $\sim t^{-3/2}$ for the excess density created by the quench there. We point out that since in our description of the collision dynamics, $G_\parallel$ adiabatically follows the evolution of $G_\perp$, the temperature quench should not couple to the transverse mode, as the adiabatic model is inappropriate for such a fast drive (remember that when including collisions the transverse mode is gapless). This explains the much lower density of transverse excitations at early times when including collisions. 

At the smaller momentum $|\mathbf{k}|\ll 1/\xi_\parallel$ collisions with longitudinal modes are the prime contribution to the occupation at late times, which, therefore, mirrors the evolution of $n_\perp$ at small momenta. Once $k_\text{Coll}\lesssim k$, the transverse mode at momentum $k$ is no longer fed by decaying transverse modes, and the collisionless decay $\sim t^{-3/2}$ takes over.

\begin{figure}[tp]
\includegraphics[width=.95\columnwidth]{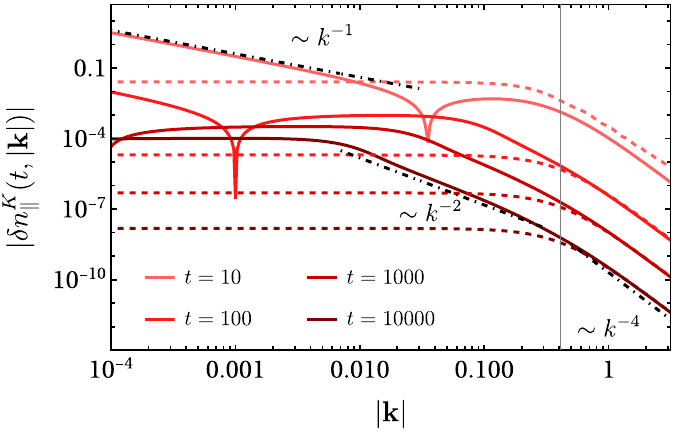}
\caption{Momentum dependence of the excess density created by the weak temperature quench \eqref{eq:quench}. Analogously to Fig.~\ref{fig:quench}, the dashed lines correspond to the collisionless approximation and solid lines to the calculation at next-to-leading order in $1/N$. The same parameters as in the main text are used $\lambda=T=\gamma=D=1$, $N=10$ and $\phi_\text{th}=\sqrt{5}$ corresponding to an inverse screening length $1/\xi\approx 0.4$, which is indicated as a gray vertical line.}
\label{fig:GKH} 
\end{figure}

The momentum dependence of $\delta n_\parallel$ at different times during the evolution is shown in Fig.~\ref{fig:GKH}. Clearly, collisions are very important at small momenta, where the gapless longitudinal modes are highly occupied by processes that involve the collision of two transverse excitations. This is to be contrasted with the evolution of gapped modes in the collisionless approximation, where the effect of the transversal excitations is limited to an effective mass. At large momenta $k\gtrsim 1/\xi_\parallel$, energy conservation prevents the scattering with the excess population in the transverse direction, and the collisionless approximation becomes faithful again. Note, that the short-time behavior at small momenta $\delta n_\parallel(\mathbf{k},t)\sim m(t) n^K_{\parallel,\text{th}}\sim t^{-5/2}\mathbf{k}^{-1}$ requires not redistribution, but a gapless spectrum. It is the direct analog of the short-time, small momentum excess $\delta n_\perp(\mathbf{k},t)\sim t^{-5/2}\mathbf{k}^{-2}$ reported in Fig.~\ref{fig:quench}. The difference in the momentum dependence is directly related to the momentum dependence of the effective decay rate.

\bibliographystyle{apsrev4-2}
\bibliography{non_thermal}

\begin{thebibliography}{106}
\bibitem [{not()}]{note:bare}%
  \BibitemOpen
  \href@noop {} {}\bibinfo {note} {To keep the notation simple we use bare
  parameters for the transverse mode. In general one has to introduce dressed
  values of $\gamma$ and $D$ in $\Pi^R_\text{th}$ etc.}\BibitemShut {Stop}%
\end{thebibliography}

\end{document}